\documentclass[aps,pra,twocolumn,superscriptaddress]{revtex4}
\usepackage[usenames,dvipsnames]{color}

\usepackage[latin1]{inputenc}
\usepackage{graphicx}
\usepackage{amssymb}
\usepackage{braket,mleftright}
\usepackage{empheq}
\usepackage{subfigure}
\usepackage{stackrel}
\usepackage{soul}
\usepackage{blkarray}
\usepackage{multirow}
\usepackage{amsmath}
\usepackage{physics}
\usepackage{amsfonts}
\usepackage{bm}
\usepackage{bbold} 
\usepackage{color}
\newcommand{\rro}{\hat{\rho}_0}
\newcommand{\Lo}{\hat L_{0}}

\newcommand{\sgn}{\ensuremath{{\mathrm{sgn}}}}

\newcommand{\beq}{\begin{equation}}
\newcommand{\eeq}{\end{equation}}
\newcommand{\beqq}{\begin{equation*}}
\newcommand{\eeqq}{\end{equation*}}
\newcommand{\bse}{\begin{subequations}}
\newcommand{\ese}{\end{subequations}}
\newcommand{\bea}{\begin{eqnarray}}
\newcommand{\eea}{\end{eqnarray}}
\newcommand{\beaa}{\begin{eqnarray*}}
\newcommand{\eeaa}{\end{eqnarray*}}
\newcommand{\ep}{\epsilon}
\newcommand{\bit}{\begin{itemize}}
\newcommand{\eit}{\end{itemize}}
\newcommand{\bpmatrix}{\begin{pmatrix}}
\newcommand{\epmatrix}{\end{pmatrix}}

\begin{document}


\title{Attainability of the quantum information bound in pure state models.}

\author{Fabricio Toscano}
\affiliation{Instituto de F\'{\i}sica, Universidade Federal do Rio de
Janeiro, Caixa Postal 68528, Rio de Janeiro, RJ 21941-972, Brazil}

\author{Wellison P. Bastos}
\affiliation{Instituto de F\'{\i}sica, Universidade Federal do Rio de
Janeiro, Caixa Postal 68528, Rio de Janeiro, RJ 21941-972, Brazil}

\author{Ruynet L. de Matos Filho }
\affiliation{Instituto de F\'{\i}sica, Universidade Federal do Rio de
Janeiro, Caixa Postal 68528, Rio de Janeiro, RJ 21941-972, Brazil}

\email[]{toscano@if.ufrj.br}

\begin{abstract}
The attainability of the quantum Cramér-Rao bound [QCR],
the ultimate limit in the precision of the estimation of a physical parameter, requires 
the saturation of the quantum information bound [QIB]. This occurs when the Fisher information associated to a  given measurement on the quantum state of a system which encodes the information about the parameter coincides with the quantum Fisher information associated to that quantum state.
Braunstein and Caves [PRL {\bf 72}, 3439 (1994)] have shown that the QIB can 
 always be achieved via a projective measurement 
in the eigenvectors basis of an observable called symmetric logarithmic derivative. However, such projective measurement depends, in general, on the value of the parameter to be estimated. Requiring, therefore, the previous knowledge  of the quantity one is trying to estimate. For this reason, it is important to investigate under which situation it is possible to saturate the QCR without previous information about the parameter to be estimated.  
Here, we show the complete solution to the problem of which are all the initial pure states 
and  the projective measurements that allow the global saturation of the QIB, without the knowledge of the true value of the parameter, when the information about the parameter is encoded in the system by a unitary process.  
\end{abstract}


\maketitle

\section{Introduction}
\label{SectionIntro}

The aim of quantum statistical estimation theory is to estimate the true
value of a real parameter $x$ through suitable 
measurements on a quantum system of interest. It is assumed that the state of the quantum system belongs to a family $\hat\rho(x)$ of density operators, defined on a Hilbert Space ${\cal H}$, and parametrized by the parameter $x$.  
The practical implementation of the estimation process comprises two steps:
the first one consists in  the acquisition of  experimental data from specifics quantum measurements on the system of interest
while the second one consists in  the data manipulation in order to obtain an estimative 
of the true value of the parameter  \cite{hayashi2005}.
The first step is implemented via  a positive-operator valued measure (POVM), described by a set of positive Hermitian operators $\{\hat E_j\}$, which add up to the identity operator ($\sum_{j=1}^N\hat E_j=\hat{\mathbb 1}$). The probability of obtaining the measurement result $j$, if the value of the parameter is $x$, is then given by $p_j(x)=\Tr[\hat\rho(x)\hat E_j] $.
The second step is implemented by using an estimator to process the data and produce an estimate 
of the true value of the parameter.
\par
It is well known that there is a fundamental limit for the minimum reachable uncertainty in the estimative of the value of a parameter $x$, produced by any estimator. When this uncertainty is quantified  by the variance  $\delta^2x$ of the estimates of $x$, this ultimate lower bound is known as the Quantum Cramér-Rao (QCR)  bound  and is given by
$\delta^2x \geq 1/\nu {\cal F}_Q(x_v)$, where ${\cal F}_Q(x_v)$ is the Quantum Fisher Information (QFI)
of the state $\hat \rho(x_v)$,  $\nu$ is the number of repetitions of the measurement on the system, and $x_v$ is the true value of the parameter.  
The  QFI is  defined as 
${\cal F}_Q(x_v)\equiv \underset{\{\hat E_j\}}{\mbox{max}}\{{\cal F}(x_v,\{\hat E_j\})\}$, where ${\cal F}(x_v,\{\hat E_j\})$ is the Fisher Information (FI) associated 
to the probabilities distributions $p_j(x_v)=\Tr[\hat \rho(x_v) \hat E_j]$.
In this regard  ${\cal F}_Q(x_v)$ is a measure
of the maximum information on the parameter $x_v$ contained in the quantum state $\hat \rho(x_v)$.
Determining the exact conditions necessary for the saturation of the fundamental limit 
of precision plays a central role  in quantum statistical estimation theory. 
\par 
Braunstein and Caves~\cite{braunstein1994} (see also \cite{braunstein1996})  have investigated and demonstrated the attainability of the QCR bound by separating it in two steps, which are represented by the two inequalities  $\delta^2x\geq 1/\nu {\cal F}(x_v,\{\hat E_j\})\geq 1/\nu {\cal F}_Q(x_v)$.
The first inequality corresponds to the Classical Cramér-Rao  (CCR) bound associated with 
the particular quantum measurement $\{\hat E_j\}$ performed on the system, where ${\cal F}(x_v,\{\hat E_j\})$ is the Fisher information
about the parameter $x_v$ associated to the set 
of probabilities  $\{p_j(x_v)\}$.
The saturation of the CCR bound depends on the nature of the estimator used
to process the data drawn from the  set of probabilities $\{p_j(x_v)\}$ in order to estimate the 
true value of the parameter. Those estimators 
that saturates the CCR bound are called {\it efficient} estimators or
{\it asymptotically efficient} estimators  \cite{rao-book} when the
saturation only occurs in the limit of a very large number $\nu$ of measured data. 
A typical example of an  {\it asymptotically efficient} estimator  is the maximum likelihood
estimator \cite{rao-book}. 
Only special families  of probabilities distributions $\{p_j(x_v)\}$ allow the construction of an
 {\it efficient} estimator for finite $\nu$. 
\par 
The second inequality applies to all quantum measurements  $\{\hat E_j\}$ and establishes  the bound 
${\cal F}(x_v,\{\hat E_j\})\leq {\cal F}_Q(x_v)$.
Saturation of this bound corresponds to finding 
{\it optimal measurements} $\{\hat E_j\}$, such that
\beq
\label{saturation-QIB}
{\cal F}(x_v,\{\hat E_j\})={\cal F}_Q(x_v).
\eeq
These are  quantum measurements that would allow one to retrieve all the information about the parameter
 encoded in the quantum state of the system. The saturation of this bound is also 
known as the saturation of the {\it Quantum Information Bound} (QIB) in quantum  statistical estimation theory
\cite{hayashi2005}.
The quest for determining the  {\it optimal measurements} for any metrological configuration 
has a long history, going back to the pioneering works of Helstrom \cite{helstrom1967} and Holevo
\cite{Holevo1982}, and has been  subject of interest of recent work~\cite{braunstein1994,fujiwara1994,braunstein1996,barndorff2000,hayashi2005}.
In order to prove the attainability of the QCR bound, the authors of Ref.~\cite{braunstein1994}  have shown that an upper bound to the QFI, based on the so-called symmetric logarithmic derivative (SLD) operator $\hat L(x)$ , was indeed equal to the QFI. 
This upper bound was first discovered by  Helstrom \cite{helstrom1967} and Holevo \cite{Holevo1982} and is given by:
\beqq
{\cal F}_Q(x_v) \le \Tr[\hat \rho(x_v)\hat L^2(x_v)].
\eeqq
The proof consists in showing that a sufficient condition
for   achieving the equalities ${\cal F}(x_v,\{\hat E_j\})={\cal F}_Q(x_v)=\Tr[\hat \rho(x_v)\hat L^2(x_v)]$
is given by the use  of  a POVM $\{\hat E_j\}$ such that the operators $\hat E_j$ are one-dimensional projection operators onto the eigenstates  of the SLD operator $\hat L(x_v)$. That is $\{\hat E_j(x_v)\}\equiv \{\ket{l_j(x_v)}\bra{l_j(x_v)}\}$, where
$\ket{l_j(x_v)}$ is an eigenstate of $\hat L(x_v)$. 
At this point  it is important to notice that although the use of this optimal POVM is sufficient to saturate the QIB, it depends, in general,  on the true value of the parameter  one  wants to estimate, {\it i.e.}  $\{\hat E_j\}=\{\hat E_j(x_v)\}$. 
\par
Mainly  two approaches have been adopted in order to deal with the fact that the optimal POVM depends on the true value $x_v$  of the parameter.
The first one relies on adaptive quantum estimation schemes that could, in principle, asymptotically achieve the QCR bound 
\cite{Fischer2000,hayashi2005ch10,hayashi2005ch13,hayashi2005ch15,fujiwara2006}.
Such  approach is valid for any arbitrary state $\hat \rho(x)$.
The second one looks for the families of density operators $\{\hat \rho(x)\}$, for which the use of   an specific 
POVM  $\{\hat E_j\}$ that does not depend on the true value of the parameter leads to the saturation of the QIB.
  Our work follows this  approach.
\par
Within the second approach, when the family $\{\hat \rho(x)\}$ corresponds to 
operators with no null eigenvalues (full rank), the analysis of the saturation of the QIB is simplified because,
given $\hat \rho(x)$, there is only one solution for the SLD operator equation:
\beq
\label{Eq-L-0perator}
\frac{d\hat \rho(x)}{dx}=\frac{1}{2}(\hat \rho(x)\hat L(x)+\hat L^\dagger(x) \hat \rho(x)),
\eeq
with $\hat L^\dagger(x)=\hat L(x)$ \cite{hayashi2005ch9}.
For  full-rank operators, Nagaoka showed~\cite{hayashi2005ch9} that  saturation of the 
quantum information bound by using a POVM that does not depend on the true value of the parameter is only possible for the so-called {\it quasi-classical  family} of density operators. He also presented complete characterisation of the quantum measurements that guarantee
the  saturation  for this family.   
Therefore, the problem of finding the states and
the corresponding optimal measurements that lead to the saturation of the QIB, independently of the true value of the parameter,
 in the case of one-parameter 
families of full-rank density operators has been  already solved.
\par
However, for the opposite case of pure states (rank-one density operators), the complete 
characterisation of the families of states  and the corresponding measurements that lead to the saturation of the QIB, independently of the true value of the parameter,  is still an open question in the case of arbitrary Hilbert spaces. It is important to remark that  inside the families of pure states   the  QFI reaches its largest values.
Among theses families,  the most important ones  are those  unitarily generated from an 
initial state $\hat \rho_0=\ket{\phi_+}\bra{\phi_+}$
as
\beq
\label{our-families}
\hat \rho(x)=e^{-i\hat A x}\;\hat \rho_0\;e^{i\hat A x},
\eeq
where the Hermitian generator $\hat A$ does not depend on the parameter $x$ to be estimated. In this case the QFI is given by~\cite{braunstein1996}:
\beq
\label{Variance}
{\cal F}_Q=4\langle(\Delta\hat{A})^{2}\rangle_{+},
\eeq
where $\langle(\Delta\hat{A})^{2}\rangle_{+}=\Tr[\hat \rho_0(\hat A-\langle \hat A\rangle_{+})^2]$ and $\langle \hat A\rangle_{+}\equiv \Tr[\hat \rho_0 \hat A]$.
\par
For these kind of families, Ref.~\cite{braunstein1996}  considered the situations where the  Hermitain operators 
$\hat A$  generate ``displacements'' on a Hilbert Space  basis  $\{\ket{x}\}$, {\it i.e.}, $e^{-i\hat A x}\ket{0}=\ket{0+x}$ , where $\ket{0}$ is an arbitrary state.  For these situations, the authors could find   all the initial states 
$\ket{\phi_+}$ and  the corresponding global optimal POVMs that saturate the  QIB, independently of the true value of the parameter $x$.
In Ref.\cite{barndorff2000} the authors investigated  under which conditions a global  saturation of QIB can happen for two level quantum systems. 
 \par
 Here, we present the complete solution to the problem of which are all the  initial states 
 $\ket{\phi_+}$ and  the corresponding
  families of global projective measurements that allow the saturation of the QIB, within the quantum state family given in Eq.(\ref{our-families}), for arbitrary  generators $\hat A$ with discrete spectrum.  We put together
a catalogue of the initial states  $\ket{\phi_+}$ that allow global saturation of the QIB according  to the number of 
eigenstates $\ket{A_{k_l}}$ ($l=1,\ldots,M$) of the generator $\hat A$ which are present in their expansion in the eigenbasis of $\hat A$.
For a fixed value of the mean $\langle \hat A\rangle_{\phi_+}$, each member of this catalogue can be expanded in terms of  a subset  $\{k_l\}$ of eigenstates $\ket{A_{k_l}}$ whose corresponding eigenvalues  are equidistant from the mean, provided  the coefficients of that expansion satisfy  certain symmetry conditions. 
We show that the global saturation of the QIB requires  specific projective measurements  within the 
subspace $\{\ket{A_{k_l}}\}_{l=1,\ldots,M}$, determined by the initial state 
$\ket{\phi_+}$, and  give the full characterisation of these projective measurements.
We also identify, among all the initial states $\ket{\phi_+}$  that lead to global saturation of  the QIB,  for a fixed value of the 
mean $\langle \hat A\rangle_{\phi_+}$, which one has the largest QFI.  When the spectrum of the generator $\hat A$ is lower bounded, such state is a balanced linear superposition of the lowest eigenstate of $\hat A$ and the eigenstate  symmetric to it in relation to the mean. 
Interestingly, the QCR bound associated to that state corresponds to the well known Heisenberg limit in quantum metrology
\cite{Giovannetti2006}. This shows that, for the situations considered in this paper, the states that lead to the Heisenberg limit saturate the QIB via projective measurements which do not depend on the true value of the parameter.
\par
The paper is organised as follows. In Section \ref{SectionII} we reformulate the conditions 
for the saturation of the QIB, first settled in \cite{braunstein1994}, in a way appropriate
to treat the one-parameter quantum state families in (\ref{our-families}). 
Next, in Section \ref{SectionIII} we find the solutions for these conditions that give the structure of all the initial states and all the projective 
measurements that  allow the saturation of the QIB without the knowledge of the true value of the parameter.  In Section \ref{SectionIV}, we applied our results in two contexts:
phase estimation in a two path-interferometry using the Schwinger representation and 
phase estimation with one bosonic mode.
Section \ref{SectionV} is devoted to show that our solutions for the saturation of the QIB include the initial states 
whose quantum Fisher information correspond to the so called Heisenberg limit
and to show that these are the initial states that allow the maximum retrieval of information about the parameter, among all initial states that saturate the QIB. 
Finally, we give in Section \ref{SectionVI} a summary of our results.

\section{Condition for  global saturation of the QIB in pure state models}
 \label{SectionII}
 
 Let's begin with an arbitrary quantum state family and consider
  the set of inequalities, first stablished in \cite{braunstein1994},
 that the Fisher information associated with a POVM $\{\hat E_j\}$ must satisfy:  
 \begin{widetext}
 \bse
 \label{condition-sat-QIB}
 \bea
\mathcal{F}(x_v,\{\hat{E}_{j}\})&=&\sum_{j}\frac{1}{\Tr[\hat{\rho}(x_v)\hat{E}_{j}]}\Bigg(\Tr\Bigg[
\left.\dv{\hat \rho (x)}{x}\right|_{x=x_v}
\hat{E}_{j}\Bigg]\Bigg)^{2}
=\sum_{j}\frac{\left(\mathrm{Re}\left(\Tr\left[\hat{\rho}(x_v)\hat{E}_{j}\hat{L}^{\dagger}
(x_v)\right]\right)\right)^{2}}{\Tr[\hat{\rho}(x_v)\hat{E}_{j}]}
\label{Desig}\\
&\leq&\sum_{j}\frac{\left|\Tr\left[\hat{\rho}(x_v)\hat{E}_{j}\hat{L}^{\dagger}(x_v)\right]\right|^{2}}{\Tr[\hat{\rho}(x_v)\hat{E}_{j}]}
=\sum_{j}\left|\Tr\left[\left(\frac{\hat{\rho}^{1/2}(x_v)\hat{E}^{1/2}_{j}}{\left(\Tr[\hat{\rho}(x_v)\hat{E}_{j}]\right)^{1/2}}\right)
\left(\hat{E}^{1/2}_{j}\hat{L}^{\dagger}(x_v)\hat{\rho}^{1/2}(x_v)\right)\right]\right|^{2}\label{Desig1}\\
&\leq&
\Tr\big[\hat{\rho}(x_v)\hat{L}(x_v)\hat{L}^{\dagger}(x_v)\big]
=\Tr\big[\hat{\rho}(x_v)\hat{L}^2(x_v)\big]
\equiv {\cal F}_Q(x_v),
\label{e4.47}
\eea
\ese
\end{widetext}
where  in Eq.(\ref{Desig}) we used  the Sylvester equation (\ref{Eq-L-0perator}), in 
Eq.(\ref{Desig1}) the inequality $\mathrm{Re}^2(z)\leq |z|^2$ ,
in Eq.(\ref{e4.47}) the Cauchy-Schwarz inequality $|\mathrm{Tr}[\hat{A}\hat{B}^{\dagger}]|^{2}\leq\mathrm{Tr}[\hat{A}\hat{A}^{\dagger}]\mathrm{Tr}[\hat{B}\hat{B}^{\dagger}]$ and the fact that $\hat{L}(x_v)$ is an Hermitian operator. The 
necessary and sufficient conditions for the saturation of the 
QIB given in (\ref{condition-sat-QIB}) can be condensed into the  requirement that the quantities
\bea
\lambda_{j}(x_v)=\frac{\mathrm{Tr}\left[\hat{\rho}(x_v)\hat{E}_{j}\hat{L}(x_v)\right]}{\mathrm{Tr}\left[\hat{\rho}(x_v)\hat{E}_{j}\right]}
\label{c3}
\eea
be real numbers for all values of $j$ and possible values of $x_{v}$.
\par
\begin{figure}[h]
\centering
\includegraphics[width=8.5cm]{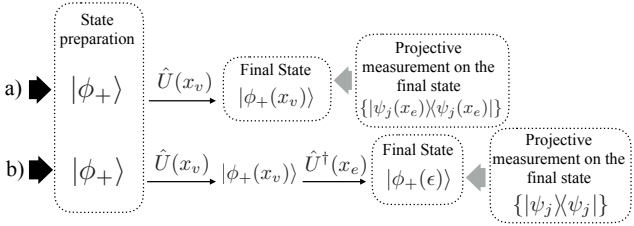}
\caption{{\bf a)} Quantum estimation process of the parameter $x_v$ corresponding to the 
laboratory's set up. In this case, the projective measurement on the final state depends on a guess value
$x_e$  for the parameter, where $\ket{\psi_j(x_e)}$ is given in Eq.(\ref{med-projetivas}). {\bf b)} Equivalent  quantum estimation process appropriate for theoretical analysis.
In this case the parameter to be estimated is $\epsilon\equiv x_v-x_e$, which is  imprinted on the 
final state given in Eq.(\ref{final-state-epsilon}), and the projective measurement on
that state  does not depend on $\epsilon$. }
\label{fig1}
\end{figure}

Let's restrict our attention to the pure quantum state family given in~(\ref{our-families}), where the generator $\hat A$  of the unitary transformation has a discrete  spectrum. In that case, if  the system is initially in the state $\ket{\phi_+}$, after the unitary transformation it will be in the state 
\beq
\label{final-state}
\ket{\phi_+(x_v)}=e^{-i\{\hat{A}-\langle\hat{A}\rangle_{+}\}x_{v}}\ket{\phi_+},
\eeq
where $x_v$ is the true value of the parameter to be estimated and the phase $e^{-i x_v \langle\hat{A}\rangle_{+}}$ guarantees that the QFI is just the variance of the generator $\hat A$ in the initial state $\ket{\phi_+}$ (see Eq.(\ref{Variance})).
We consider now projective quantum measurements on the system, described by the projectors 
\bea
\hat{E}_{j}(x_{e})&=&\dyad{\psi_j(x_{e})}=\nonumber\\
&=&e^{-i\{\hat{A}-\langle\hat{A}\rangle_{+}\}x_{e}}
\dyad{\psi_j}e^{i\{\hat{A}-\langle\hat{A}\rangle_{+}\}x_{e}},
\label{med-projetivas}
\eea
which may depend on a guess  $x_e$ at the true value of the parameter, based, for example, on some prior information about that value. 
Here,
$\{\ket{\psi_j}\}$ is a countable basis of the Hilbert space of the system. 
 The probability of getting the result $j$ in the projective measurement
$\{\dyad{\psi_j(x_{e})}\}$ can then be written as
 \beaa
&p_{j}(x_{e},x_v)=\mathrm{Tr}[\dyad{\phi_+(x_v)}\hat{E}_{j}(x_{e})]=\nonumber\\
&=\mathrm{Tr}\big[\dyad{\phi_+(\epsilon)}\,\dyad{\psi_j}\big]=p_{j}(\epsilon),
\eeaa
where  we define 
\beq
\epsilon=x_v-x_e.
\label{def-epsilon}
\eeq 
Notice that  $p_j(\epsilon)$ corresponds equivalently to the probability  of getting the result  $j$ in the projective measurement
$\{\dyad{\psi_j}\}$ on the final state
\beq
\label{final-state-epsilon}
\ket{\phi_+(\epsilon)}=\hat U(\epsilon)\ket{\phi_+}
=e^{-i\{\hat{A}-\langle\hat{A}\rangle_{+}\}\epsilon}\ket{\phi_+}.
\eeq
The relation between the Fisher information associated to the measurement
$\{\dyad{\psi_j(x_{e})}\}$ on the state $\ket{\phi_+(x_v)}$ and 
the Fisher information associated to the measurement
$\{\dyad{\psi_j}\}$ on the state $\ket{\phi_+(\epsilon)}$ is
\beqq
\mathcal{F}(x_v,\{\ket{\psi_j(x_e)}\})=\mathcal{F}(\epsilon, \{\ket{\psi_j}\})\equiv \mathcal{F}(\epsilon),
\eeqq
where we use $\pdv*{p_j(x_e,x)}{x}|_{x=x_v}=\dv*{p_j(\epsilon^\prime)}{\epsilon^\prime}|_{\epsilon^\prime=\epsilon}$.
Therefore, the estimation of the true value $x_v$ of the parameter $x$ in the pure state family given in Eq.(\ref{final-state}) via the projective measurement $\{\dyad{\psi_j(x_e)}\}$, which 
depends on the guess  value $x_e$, is equivalent to the estimation of the parameter $\epsilon$ 
in  the pure state family given in Eq.(\ref{final-state-epsilon}) 
via the projective measurement $\{\dyad{\psi_j}\}$, which does not depend on the values 
$x_e$ and  $x_v$ (see Fig.(\ref{fig1})).
\par
The Sylvester equation that define the SLD operator associated with the states $\ket{\phi_+(\epsilon)}$ can be written as an algebraic equation between operators:
\bea
\hat L^\prime_0&=&2 \hat U^\dagger(\epsilon)\left.\dv{\hat \rho(\epsilon^\prime)}{\epsilon^\prime}\right|_{\epsilon^\prime=\epsilon}
 \hat U(\epsilon)\nonumber\\
 &=&
 \rro \Lo(\epsilon)+\Lo(\epsilon)\hat \rho_0, 
 \label{SLD-equation-Lo}
\eea
 where $\hat \rho_0\equiv \dyad{\phi_+}$, and
 \beaa
 \hat L^{\prime}_0&\equiv& 2i[\rro, (\hat A-\langle \hat A\rangle_{+})],
 \\
 \Lo(\epsilon)&\equiv& \hat{U}^\dagger(\epsilon) \hat L(\epsilon)\hat{U}(\epsilon).
 \eeaa
 Given an initial state $\rro$, the structure of the infinite solutions $\Lo(\epsilon)$ of  Eq.(\ref{SLD-equation-Lo}) can be better displayed  if one defines the auxiliary state
 \bea
|\phi_{-}\rangle\equiv\frac{-2i}{\sqrt{\mathcal{F}_Q}}(\hat{A}-\langle\hat{A}\rangle_{+})\ket{\phi_{+}},
\label{estortogonal}
\eea
orthogonal to the initial state $\ket{\phi_+}$. 
In this case, one  can rewrite the 
operator $\Lo^{\prime}$ as
\begin{equation*}
\Lo^{\prime}= 
\sqrt{\mathcal{F}_Q}\big(\ket{\phi_+}\bra{\phi_-}+\ket{\phi_-}\bra{\phi_+}\big),
\end{equation*}
with $\mathcal{F}_Q=4\langle(\Delta\hat{A})^{2}\rangle_{+}$.
Let's introduce now a countable basis $\{\ket{\phi_k}\}$ of the Hilbert space of the system, 
 with $\ket{\phi_1}=\ket{\phi_+}$ and $\ket{\phi_2}=\ket{\phi_-}$.
In this basis, 
all the solutions $\Lo(\epsilon)$ have the matrix structure
\begin{equation}
\label{mat-L0-ep}
\begin{blockarray}{ccccccc}
 & |\phi_{+}\rangle & |\phi_{-}\rangle & |\phi_{3}\rangle & \cdots & |\phi_{k}\rangle&\cdots  \\
\begin{block}{c(c|ccccc@{\hspace*{5pt}})}
 \langle\phi_{+}|  &     0      & \sqrt{\mathcal{F}_Q} & 0 & \cdots & 0 & \cdots\\ 
  \cline{2-7}
 \langle\phi_{-}|  & \sqrt{\mathcal{F}_Q} &          & &  &  & \\
 \langle\phi_{3}|  &      0       &    &  \BAmulticolumn{4}{c}{\multirow{4}{*}{$\mathbb{L}(\ep)$}}\\
 \vdots            &   \vdots     &     &\\
  \langle\phi_{k}|              & 0       &      &\\
\vdots&\vdots            &  &\\ 
\end{block}
\end{blockarray}\;,
\end{equation}
where $\mathbb{L}(\epsilon)$ is an arbitrary Hermitian matrix. When the matrix $\mathbb{L}(\epsilon)$ is the null matrix we recover the particular solution $\Lo^\prime$. 
We stress that Eq.(\ref{SLD-equation-Lo}) has an infinite number of solutions even if
$\epsilon=0$ ({\it i.e.} when the guess value  $x_e$ coincides with the true
value $x_v$) because $\mathbb{L}(0)$ is not necessarily the null matrix. 
\par
We are now able to rewrite the  saturation conditions of the QIB in Eq.(\ref{c3}) for our 
pure quantum state family models as the requirement that
\bse
\label{c1-lambda}
\bea
\lambda_{j}(\epsilon)&=&
\frac{\mathrm{Tr}\left[\hat{\rho}(\epsilon)\,\dyad{\psi_j}\,\hat{L}(\epsilon)\right]}{\mathrm{Tr}\left[\hat{\rho}(\epsilon)\,\dyad{\psi_j}\right]}=\label{c1-lambda-1}\\
&=&
\frac{\matrixel{\psi_j}{\hat U(\epsilon)\hat{L}_{0}(\epsilon)}{\phi_+}}
{\braket{\psi_j}{\phi_+(\ep)}}=
\label{c1-lambda-2}\\
&=&\sqrt{\mathcal F_Q}\frac{\bra{\psi_j}\phi_-(\ep)\rangle}{\bra{\psi_j}\phi_+(\ep)\rangle}
\label{c1-lambda-3}
\eea
\ese
be real numbers. Here we define
\beq
\label{state-phi-minus-epsilon}
\ket{\phi_-(\epsilon)}=\hat U(\epsilon)\ket{\phi_-}
=e^{-i\{\hat{A}-\langle\hat{A}\rangle_{+}\}\epsilon}\ket{\phi_-}.
\eeq
In Eq.(\ref{c1-lambda-2}), we used the fact that, according to Eq.~(\ref{mat-L0-ep}), all SLD operators $\Lo(\epsilon)$ verify
$\Lo(\ep)\ket{\phi_+}=\Lo^{\prime}\ket{\phi_+}=\sqrt{F_Q}\ket{\phi_-}$
for all values of $\epsilon$.
\par
From Eq.(\ref{c1-lambda-2}) one can see that,  when $\epsilon=0$ ({$x_e=x_v$),
if the states $\ket{\psi_j}$ are eigenstates of $\hat L_0(0)$, then the conditions in Eqs.(\ref{c1-lambda})
are automatically satisfied for all values of $j$
and we recover in our formalism the conditions for the saturation of the QIB first stated
in \cite{braunstein1994}. We are, however, interested in finding the conditions  for global saturation of the
QIB, which correspond to all the initial states $\ket{\phi_+}$ and all the projective measurements 
$\{\dyad{\psi_j}\}$ that allow the saturation of the QIB for all values of $\epsilon$. 
This is equivalent to finding projective measurements on the final state
$\ket{\phi_+(x_v)}$
 that, regardless of the true value $x_v$ of the parameter, lead to the saturation of the QIB.
For this sake, it is convenient to rewrite the inequalities that must be satisfied by the Fisher information  $\mathcal{F}(\ep)$ as 
\beaa
\mathcal{F}(\ep)&=&
\mathcal{F}_Q\left(1-
\sum_{j}\frac{\left(\mathrm{Im}
\big[w_j(\epsilon)z^*_j(\epsilon)]\right)^{2}}{p_j(\epsilon)}\right)\nonumber\\
&\leq &\mathcal{F}_Q=4\langle(\Delta\hat{A})^{2}\rangle_{\phi_+},
\eeaa
where  
\bse
\label{def-zj-wj}
\bea
\braket{\psi_j}{\phi_+(\epsilon)}
&\equiv& | z_{j}(\epsilon)| e^{i\alpha_{j,+}(\epsilon)} ,\\  
 \braket{\psi_j}{\phi_-(\epsilon)}
&\equiv& w_{j}(\epsilon),
\eea
\ese
with $\alpha_{j,+}(\epsilon)=\arg(z_{j}(\epsilon))$.
\par
This yields conditions which are equivalent to those in Eq.(\ref{c1-lambda}) and can be written as
\beq
\mathrm{Im}\big[w_{j}(\epsilon)z^{*}_{j}(\epsilon)\big]=0.
\label{cond-sat-Im}
\eeq
 Notice that the relation above must apply for any value of $\epsilon$ and  for all $j$.
For future use we rewrite the conditions in Eq.(\ref{cond-sat-Im})
as:
\bea
&\sum_{\substack{j' \neq j}}|v_{j,j'}|
\frac{|z_{j'}(\ep)|}{|z_{j}(\ep)|}\cos\left(\alpha_{j',+}(\ep)-\alpha_{j,+}(\ep)+\phi_{j,j'}\right)=\nonumber\\
&=\langle\hat{A}\rangle_{+}-v_{j,j},
\label{sat-psi-j2}
\eea
where we define 
$\bra{\psi_j}\hat{A}\ket{\psi_{j^\prime}}\equiv|v_{j,j^\prime}|e^{i\phi_{j,j^\prime}}$.
\par
\section{Projective measurements and states for a global
saturation of the QIB}
\label{SectionIII}

Any initial state $\ket{\phi_+}$ of the quantum state family given in Eq.(\ref{final-state-epsilon})
can be written in the basis of eigenstates of the generator $\hat A$. 
In order to find the initial states that allow global saturation of the QIB, we
consider states $\ket{\phi_+}$ that are finite linear combinations of the eigenstates 
 $\ket{A_{k_l}}$ of $\hat A$:
\bea
\ket{\phi_+}=\sum_{l=1}^{M}|c_{k_l}|\,e^{i\theta_{k_l}}\ket{A_{k_l}},
\label{phi-base-ger}
\eea
with $\theta_{k_l}=\arg(c_{k_l})$ and $c_{k_l}\neq 0$.
The set of integers $\{k_l\}_{l=1,\ldots,M}$ with $k_1 < k_2 < \ldots < k_M$  are the labels of the eigenstates that 
define the subspace $\{\ket{A_{k_l}}\}_{l=1,\ldots,M}$ of the 
Hilbert space of the system. 
When  the spectrum of $\hat A$ is unbounded, the initial states $\ket{\phi_+}$ may have an infinite number of terms in an expansion like in Eq.(\ref{phi-base-ger}). In this case, as it will be shown, depending on the class of projective measurements one uses, one  either  takes the limit  $M\rightarrow \infty$ or have to consider instead approximated states, which  correspond to a truncation up to sufficiently large $M$ terms in Eq.(\ref{phi-base-ger}).
It is also important to notice that the evolved state $\ket{\phi_+(\epsilon)}$ remains in the 
subspace $\{\ket{A_{k_l}}\}_{l=1,\ldots,M}$ for all values of $\epsilon$.
We also assume that the mean value $\langle \hat A\rangle_{+}$ is a predetermined fixed quantity
and therefore all the considered initial states  have to satisfy this constraint.
\par
A 
global saturation of the QIB for a initial state
$\ket{\phi_+}$ and a projective measurement $\{\dyad{\psi_j}\}$  means that 
\bse
\label{eq20}
\bea
\mathcal{F}(\ep)&=&\sum_jp_j(\epsilon)\lambda_j^2(\epsilon)=
\label{eq20a}\\
&=&\mathcal F_Q\sum_{j}\braket{\phi_+(\epsilon)}{\psi_j}
\frac{\braket{\psi_j}{\phi_-(\epsilon)}^2}{\braket{\psi_j}{\phi_+(\epsilon)}}=
\label{eq20b}
\\
&=&\mathcal F_Q
\matrixel{\phi_-(\epsilon)}{\left(
\sum_{j} \dyad{\psi_j}{\psi_j}\right)}{\phi_-(\epsilon)}=
\label{eq20c}\\
&=&\mathcal{F}_Q,
\eea
\ese
 for all values of $\epsilon$.
 From Eq.(\ref{eq20a}) to (\ref{eq20b}) we use the definition of
 $\lambda_{j}(\ep)$ given in Eq.(\ref{c1-lambda}) and from (\ref{eq20b}) to (\ref{eq20c})
 we use that $\lambda_{j}^*(\ep)=\lambda_{j}(\ep)$. 
Therefore the last equality in Eqs.(\ref{eq20}) holds only if projectors 
 $\{\dyad{\psi_j}\}$ span the  subspace 
 wherein the evolved state $\ket{\phi_+(\epsilon)}$ lives,
 {\it i.e.}
 \beq
\label{complete-cond}
\hat{\Bbb1}_{M}\equiv\sum_{l=1}^{M}\ket{A_{k_l}}\bra{A_{k_l}}=\sum_{j=1}^{M}\ket{\psi_j}\bra{\psi_j},
\eeq
where we  used the fact that the projectors  $\{\dyad{\psi_j}\}$ are linearly independent.
For this reason, one can write
\beq
\ket{\psi_j}=\sum_{l=1}^{M} |b_{j,k_l}|\,e^{i\theta_{j,k_l}}\;\ket{A_{k_l}},
\label{psij-na-base-de-A}
\eeq 
where $\theta_{j,k_l}=\arg(b_{j,k_l})$.
\par
Now, using the expansions in Eqs. (\ref{phi-base-ger}) and (\ref{psij-na-base-de-A}), and the definition 
of the state $\ket{\phi_-}$ in (\ref{estortogonal}),  we arrive to
\bse
\label{def-w-and-z}
\bea
z_{j}(\epsilon)& =&\sum_{l=1}^{M}  |c_{k_l}|\; |b_{j,k_l}|\times \nonumber\\
&&\times e^{i\{-(A_{k_l}-\langle\hat{A}\rangle_{\phi_+})\epsilon-\theta_{j,k_l}+\theta_{k_l}\}},
\label{z}\\
w_{j}(\epsilon)&=&\frac{-2i}{\sqrt{\mathcal{F}_Q}}\sum_{l=1}^{M} |{c}_{k_l}|\;|b_{j,k_l}|\;(A_{k_l}-\langle\hat{A}\rangle)\times\nonumber\\
&&\times e^{i\{-(A_{k_l}-\langle\hat{A}\rangle_{\phi_+})\epsilon-\theta_{j,k_l}+\theta_{k_l}\}}
\label{w}.
\eea
\ese
In order to obtain the structure of the initials states $\ket{\phi_+}$ 
and the projective measurements $\{\dyad{\psi_j}\}_{j=1,\ldots,M}$ that allow 
for a global saturation of the QIB, we substitute  Eqs.(\ref{def-w-and-z}) in Eqs.(\ref{cond-sat-Im}) and analyse which are the conditions that the sets $\{A_{k_l}\}$,
$\{c_{k_l}\}$ and $\{b_{j,k_l}\}$ with $j,l=1,\ldots,M$ must satisfy in order 
to be  solutions of these equations. 
This is done in the following section.
\\
\subsection{Structure of the initial states and the projective measurements}
\label{SubSectionIIIa}
\par
 In Appendix \ref{AppendixA}, we show that if the set of eigenstates $\{\ket{A_{k_l}}\}_{l=1,\ldots,M}$  present in the decomposition  of $\ket{\phi_+}$ does not contain two eigenstates $\ket{A_{k_l}}$ corresponding to the same eigenvalue of $\hat A$,
Eqs.(\ref{cond-sat-Im}) are satisfied if and only if the sets $\{A_{k_l}\}$,
$\{c_{k_l}\}$ and $\{b_{j,k_l}\}$ with $j,l=1,\ldots,M$, verify the conditions
\bse
\label{cond-ea}
\bea
A_{k_l}-\langle\hat{A}\rangle_{+}&=-(A_{k_{\delta(l)}}-\langle\hat{A}\rangle_{+}),\label{ea1}\\
|c_{k_l}||b_{j,k_l}|&=|c_{k_{\delta(l)}}||b_{j,k_{\delta(l)}}|,
\label{ea2}\\
(\theta_{k_{\delta(l)}}-\theta_{j,k_{\delta(l)}})&+(\theta_{k_l}-\theta_{j,k_l})=\xi_{j},
\label{ea3}
\eea
\ese 
where $\xi_j$ are arbitrary real numbers.
When $\xi_j=n_{j}\pi$, where $n_{j}$  is an integer, the solutions correspond to 
 real wave functions  $z_j(\epsilon)=\braket{\psi_j}{\phi_+(\epsilon)}$ and 
$w_j(\epsilon)=\braket{\psi_j}{\phi_-(\epsilon)}$, and when 
$n_j$ is odd, the solutions correspond to pure imaginary wave functions.
Here, $\delta(l)\equiv M-(l-1)$, for $l=1,2,\cdots,\lceil M/2\rceil$, where $\lceil \ldots \rceil$ is the ceiling function. 
It is interesting to note that when $M=2$, Eq.(\ref{ea3}) does not constitute a restriction 
on the two phases, $\theta_{k_1}$ and $\theta_{k_2}$, which appear in the expansion of the initial state 
$\ket{\phi_+}$ in the Eq.(\ref{phi-base-ger}). In this case, 
using only the conditions in Eqs.(\ref{ea1}) and (\ref{ea2}), one can show that $w_j(\epsilon)z_j^*(\epsilon)$ is given by
\bea
\label{w-zc-M-2}
&w_j(\epsilon)z_j^*(\epsilon)=4|c_{k_1}|^2|b_{k_2}|^2(A_{k_1}-\langle\hat A\rangle_+)\times\nonumber\\
&\times\sin\left(2(A_{k_1}-\langle\hat A\rangle_+)\epsilon+ 
\theta_{k_2}-\theta_{k_1}+\theta_{j,k_1}-\theta_{j,k_2}\right),
\eea
that it is always real. Therefore,  the condition of the saturation of the QIB in 
Eq.(\ref{cond-sat-Im}) is fulfilled independently of the values of the phases 
$\theta_{k_1}$ and $\theta_{k_2}$. 
Notice also that $w_j(\epsilon)z_j^*(\epsilon)$, in Eq.(\ref{w-zc-M-2}) [$j=1,2$], is real 
independently of the values  $\theta_{j,k_1}$ and $\theta_{j,k_2}$ of the phases that appear in the expansion 
of the states $\ket{\psi_j}$ of the projective measurement basis in Eq.(\ref{psij-na-base-de-A}).
This means, in particular, that if  an initial state 
$\ket{\phi_+}$  saturates the QIB with a projective measurement basis 
$\{\ket{\psi_j}\}_{j=1,2}$, then it also saturates the QIB 
  with any projective measurement basis 
 $\{|{\tilde\psi_j}\rangle=e^{ih(\hat A)}\ket{\psi_j}\}_{j=1,2}$ ,where  $h(\hat A)$ is  real function of the operator $\hat A$.
 \par
  When $M>2$,  Eq.(\ref{ea3}) fixes the relations between the  phases 
 $\theta_{k_l}$ and $\theta_{k_{\delta(l)}}$, of the
 initial state $\ket{\phi_+}$, and the phases $\theta_{j,k_l}$ and $\theta_{j,k_{\delta(l)}}$, 
 of the states $\ket{\psi_j}$ of the projective measurement basis, that indeed are crucial for 
 the saturation of the QIB. 
\par
If there are some eigenstates $\ket{A_{k_l}}$ in the decomposition of 
$\ket{\phi_+}$ corresponding to the same eigenvalue of $\hat A$, then Eqs.(\ref{cond-ea}) are only sufficient conditions 
to get  equality in Eqs.(\ref{cond-sat-Im}). However, in this case we can not guarantee that they are also necessary conditions. 
\par
Inserting the conditions given in 
 Eqs.(\ref{cond-ea}) into  the expression for $z_{j}(\epsilon)$, given  in Eqs. (\ref{z}), one gets
 \beq
\label{zj}
z_{j}(\epsilon)= e^{i\left(\frac{\xi_j}{2}+s_j(\epsilon)\pi\right)}| \eta^{(M)}_{j}(\epsilon)|,
\eeq
with the integer $s_j(\epsilon)$ defined as 
$e^{is_j(\epsilon)\pi}=\sgn(\eta^{(M)}_{j}(\epsilon))$ 
and where we also define
\begin{widetext}
\beq
\eta^{(M)}_{j}(\epsilon)=
\left\{
   \begin{matrix} 
  2\sum^{\lceil M/2\rceil}_{l=1}|c_{k_{l}}||b_{j,k_{l}}| \cos\left((A_{k_{l}}-\langle\hat{A}\rangle_+)\epsilon+(\theta_{k_{l}}-\theta_{j,k_{l}})\right) ,  & \mbox{for even $M$}, \\
  2\sum^{\lceil M/2\rceil-1}_{l=1}|c_{k_{l}}||b_{j,k_{l}}| \cos\left((A_{k_{l}}-\langle\hat{A}\rangle_+)\epsilon+(\theta_{k_{l}}-\theta_{j,k_{l}})\right) +
  \left|c_{k_{\lceil M/2\rceil}}\right| \left|b_{j,k_{\lceil M/2\rceil}}\right|& \mbox{for odd $M$}.\\
   \end{matrix}
\right..
\eeq
\end{widetext}
Therefore, the phase of  the wave function $z_{j}(\epsilon)=\braket{\psi_j}{\phi_+(\epsilon)}$ is
\beq
\label{fase-alpha}
\alpha_{j,+}(\epsilon)=\arg(z_j(\epsilon))=\left(\frac{\xi_j}{2}+s_j(\epsilon)\pi\right),
\eeq 
for all values of 
$\epsilon$. 
%
\subsection{Interpretations of the conditions for a global saturation of the QIB}
\label{SubSectionIIIb}

The condition given in Eq.(\ref{ea1}) establishes the symmetry that the 
subsets of eigenvalues $\{A_{k_l}\}_{l=1,\ldots,M}$ of the generator $\hat A$, whose respective eigenstates enter in the decomposition of the initial state $\ket{\phi_+}$, must exhibit.
This symmetry is sketched in Fig.(\ref{fig2}). It requires that, given a fixed value for 
$ \langle\hat A\rangle_{+}$, the expansion of the initial state $\ket{\phi_+}$ in the eigenbasis of $\hat A$ contains $\lceil M/2 \rceil$($M$ even) or $\lceil M/2 \rceil-1$($M$ odd)  pairs of eigenstates of $\hat A$, each pair corresponding to symmetric eigenvalues, $A_{k_l}$ and $A_{k_{\delta(l)}}$, with respect to  the mean $ \langle\hat A\rangle_{+}$. Notice that, for an arbitrary generator $\hat A$, such a expansion with $M>2$ may not exist. This is not the case if the spectrum of $\hat A$ is equally spaced. On the other hand, it is always possible to find initial states $\ket{\phi_+}$ whose expansion in the eigenbasis of $\hat A$ contains $M=2$ eigenstates and that satisfy condition (\ref{ea1}) for arbitrary generators $\hat A$.
  
\par
If we  now use in Eq.(\ref{ea2}) the orthonormality of the measurement basis vectors $\{\ket{\psi_j}\}$
\beq
\label{complete-cond2}
\sum_{j=1}^M b_{j,k_l}b^*_{j,k_{l^\prime}}=\delta_{ll^\prime},
\eeq
we get conditions for the moduli of the expansion coefficients 
of the initial state  $\ket{\phi_+}$ and of the states  $\ket{\psi_j}$ of the measurement basis in terms of the eigenstates of the generator $\hat A$:
\beq
|c_{k_l}|=|c_{k_{\delta(l)}}|,\;\; (l=1,\ldots,M)
\label{cond-phi-psi-sat1}
\eeq
and 
\beq
|b_{j,k_l}|=|b_{j,k_{\delta(l)}}|,\;\;(j,l=1,\ldots,M),
\label{balance-cond-bases}
\eeq
respectively.
These conditions imply that the eigenstates $\ket{A_{k_l}}$ and $\ket{A_{k_{\delta(l)}}}$ appear with equal weights  in the expansion of the initial state and of the measurement basis states  in terms of the eigenbasis of $\hat A$.
They also imply that, in order to allow a saturation of the QIB for all values of $\epsilon$, both the initial state $\ket{\phi_+}$ and the states
$\ket{\psi_j}$ of the projective measurement basis must have zero {\it skewness} relative to the operator $\hat A$. For example, 
it is straightforward to verify that, for the initial state 
$\hat \rho_0\equiv\dyad{\phi_+}{\phi_+}$ , the condition
in Eq.(\ref{cond-phi-psi-sat1}) leads to
\bea
S&\equiv&\mathrm{Tr}\left[\hat \rho_0 \left(\frac{\hat A-\langle \hat A \rangle_+}{\sqrt{\langle \Delta^2 \hat A \rangle}}\right)^3\right]=\nonumber\\
&=&\frac{\langle\hat A^3\rangle-\langle\hat A\rangle^3-3\langle\hat A\rangle
\langle \Delta^2\hat A\rangle}{(\langle \Delta^2\hat A\rangle)^{3/2}}=0,
\label{def-skewness}
\eea
where $S$ is the {\it skewness} of the state $\hat \rho_0$ relative to the generator $\hat A$. 
\par
Using Eq.(\ref{ea1}), we see that
\beq
\langle\hat{A}\rangle_{+}=\frac{A_{k_l}+A_{k_{\delta(l)}}}{2},\;\;\mbox{for all $l,\ldots,M$}.
\label{media-bacana}
\eeq
It is easy to check that, for all the initial states $\ket{\phi_+}$ in Eq.(\ref{phi-base-ger}) with 
$A_{k_l}$ verifying the symmetry in Eq.(\ref{ea1}) and also   
the balance condition in Eq.(\ref{cond-phi-psi-sat1}), the mean value of the generator $\hat A$ coincides with the pre-fixed value $\langle\hat{A}\rangle_{+}$:
\begin{widetext}
\beq
\bra{\phi_+}\hat{A}\ket{\phi_+}\equiv\sum_{l=1}^{M} |c_{k_l}|^2\;A_{k_l}=
   \left\{\begin{matrix} 
      \sum_{l=1}^{\lceil M/2\rceil-1} 2\,|c_{k_l}|^2\;\frac{A_{k_l}+A_{k_{\delta(l)}}}{2}+
      |c_{\lceil M/2\rceil}|^2 A_{k_{\lceil M/2\rceil}} &\;, \mbox{if $M$ is odd } \\
        \sum_{l=1}^{M/2} 2\,|c_{k_l}|^2\;\frac{A_{k_l}+A_{k_{\delta(l)}}}{2} &\;, \mbox{if $M$ is even }  \\
   \end{matrix}\right\}=
\langle\hat{A}\rangle_{\phi_+},
\label{check-mean-A}
\eeq
\end{widetext}
where we used  Eq.~(\ref{media-bacana}), the normalization condition, $\sum_{l=1}^{M}|c_{k_l}|^2=1$, of the state $\ket{\phi_+}$, and, if $M$ is odd, that $\langle\hat A\rangle_{+}=A_{\lceil M/2\rceil}$.
\par
We can also check that all the states $\{\ket{\psi_j}\}$ of a projective measurement basis that satisfy the balance condition in Eq.(\ref{balance-cond-bases}) and the condition on the phases
in Eq.(\ref{ea3}), satisfy the conditions for the a global saturation of the QIB, given in
Eq.(\ref{sat-psi-j2}). 
Indeed, 
 using (\ref{complete-cond}), 
(\ref{ea3}) and (\ref{balance-cond-bases})  we get, for $j\neq j^\prime$,
\bea
&|v_{j,j'}|e^{i\phi_{j,j^\prime}}\equiv\bra{\psi_{j}}\hat{A}\ket{\psi_{j^\prime}}=
\bra{\psi_{j}}(\hat{A}-\langle \hat A\rangle_+ )\ket{\psi_{j^\prime}}
\nonumber\\
&=2\;e^{i\left(\frac{\xi_j-\xi_{j^\prime}}{2}+\frac{\pi}{2}\right)}\sum_{l=1}^{\lceil M/2\rceil}|b_{j,k_{l}}||b_{j^{\prime},k_{l}}|\times\nonumber\\
&\times(A_{k_{l}}-\langle\hat{A}\rangle)\sin\left(\theta_{j^{\prime},k_{l}}-\theta_{j,k_{l}}-
(\xi_j-\xi_{j^\prime})/2\right)\equiv\nonumber\\
&\equiv2\;e^{i\left(\frac{\xi_j-\xi_{j^\prime}}{2}+\frac{\pi}{2}+s^\prime_{j,j^\prime}\pi\right)}|\eta^{\prime}_{j,j^\prime}|,
\eea
with $e^{is^\prime_{j,j^\prime}\pi}=\sgn(\eta^{\prime}_{j,j^\prime})$,  so that
\beq
\phi_{j,j^\prime}=\pi/2+(\xi_j-\xi_{j^\prime})/2+s^\prime_{j,j^\prime}\pi.
\label{phijjprime}
\eeq
Gathering together the results in Eqs.(\ref{fase-alpha}) and (\ref{phijjprime}) we 
obtain:
\beqq
\alpha_{j^\prime,+}(\ep)-\alpha_{j,+}(\ep)+\phi_{j,j^\prime}=
\frac{\pi}{2}+(s_{j^\prime}(\epsilon) -
s_j(\epsilon)+s^\prime_{j,j^\prime})\pi.
\eeqq
If we insert the above relation into the saturation condition of Eq.(\ref{sat-psi-j2}),
the  l.h.s of that equation turns  equal to zero. On the other hand, 
in an analogous way to that used in Eq.(\ref{check-mean-A}), we can show
that 
\beq
\bra{\psi_j}\hat{A}\ket{\psi_{j}}=\langle\hat{A}\rangle_{+} \quad\mbox{for}\quad  j=1,\ldots,M. \label{eq-meanA}
\eeq
Therefore,
 the r.hs. of Eq.(\ref{sat-psi-j2}) is also null by virtue of the balance 
condition on the coefficients in Eq.(\ref{balance-cond-bases}) and the symmetry 
of  the spectrum $\{A_{k_l}\}_{l=1,\ldots,M}$, given in (\ref{ea1}).
\begin{widetext}
\begin{figure*}[h]
\centering
\includegraphics[width=16cm]{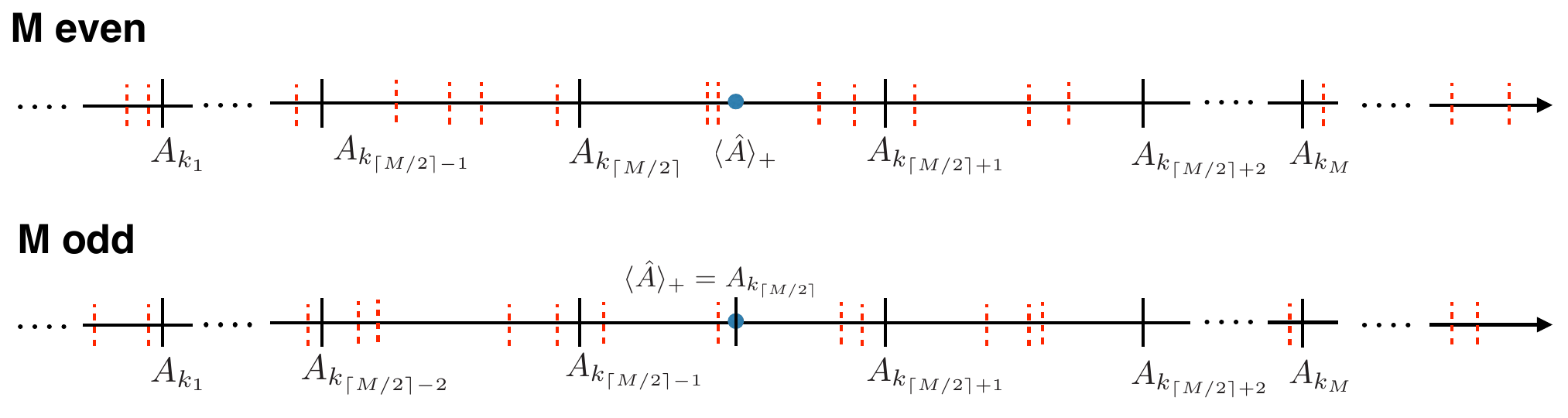}
\caption{
The vertical dashed(red) and full(black) lines represent the position of the 
eigenvalues of the Hermitian generator $\hat A$ along the real line.
The blue dot indicates the location of the fixed  mean value
$\langle\hat{A}\rangle_{+}$.
The full(black) vertical lines correspond to the subset of eigenvalues $\{A_{k_l}\}_{l=1,\ldots,M}$, whose corresponding eigenstates were used to construct the initial state $\ket{\phi_+}$  and, therefore, verify the symmetry  
in Eq.(\ref{ea1}). 
The dashed(red) vertical lines correspond to the rest of the spectrum of $\hat A$
that do not enter in the construction of the initial state $\ket{\phi_+}$.}
\label{fig2}
\end{figure*}
\end{widetext}
\par
\subsection{Projective measurements for a global saturation of the QIB}
\label{SubSectionIIIc}
\par
In the previous section, we have shown that the states of a projective measurement basis 
$\{\ket{\psi_j}\}_{j=1,\ldots,M}$ that leads to a  global saturation of the QIB 
must have  a balanced decomposition 
in terms of the subset  $\{\ket{A_{k_l}}\}_{l=1,\ldots,M}$ of  eigenstates of the generator $\hat A$. That is, the coefficients $b_{j,k_l}=\braket{A_{k_l}}{\psi_j}$  of the decomposition must verify the conditions in
(\ref{balance-cond-bases}) and (\ref{ea3}).
However, the orthonormality of the measurement basis states $\ket{\psi_j}$ places supplementary
conditions on the coefficients $b_{j,k_l}$.
In what follows we will show two examples of families of projective measurements that fulfill all the requirements for allowing a global saturation of the QIB. 
\par
\subsubsection{First family of projective measurements}
\par
We arrive at the  first family of projective measurements when,  based on Eq.(\ref{eq-meanA}),  we investigate the structure of  the measurement basis  $\{\ket{\psi_j}\}$ that satisfies the condition  $\bra{\psi_j}\hat{A}\ket{\psi_j}=\alpha$ ($j=1,\ldots,M$), where
the constant $\alpha$ does not depend on the value of $j$ and  is not necessarily equal to 
$\langle\hat{A}\rangle_{+}$.  
In Appendix \ref{AppendixB} we show that one solution to this condition corresponds to a 
decomposition of the states $\ket{\psi_j}$ in terms of the eigenstates 
$\{\ket{A_{k_l}}\}_{l=1,\ldots,M}$ with 
coefficients:
\bea
b_{j,k_l}=\frac{1}{\sqrt{M}}e^{i\theta_{j,k_l}},
\label{mod-psi-j}
\eea
where the phases are
\beq
\theta_{j,k_l}=  (j\pi/M)f_l+j\beta/M+ \phi_{k_l},
\label{fase-psi-j}
\eeq
with
\beq
\label{fm}
f_{l}=
\left\{
   \begin{matrix} 
      (l-1)+[(-1)^{l}+1](M-1)/2,  & \mbox{for even $M$}, \\
      (l-1)(1-M),   & \mbox{for odd $M$},\\
   \end{matrix}
\right.
\eeq
and  $\beta$ and $\phi_{k_l}$  arbitrary real numbers.
\par
Now, when $M>2$, using Eq.(\ref{fase-psi-j}) in Eq.(\ref{ea3}), we get for the phases of the initial state
$\ket{\phi_+}$: 
\beaa
\label{fm2}
&\theta_{k_l}+\theta_{k_{\delta(l)}}=\nonumber\\
&\left\{
   \begin{matrix} 
     \frac{j}{M}(2\pi(M-1)+2\beta)+\xi_j+\phi_{k_l}+\phi_{k_{\delta(l)}},& \mbox{$M$ even}, \\
     \frac{j}{M}(-\pi(M-1)^2+2\beta)+\xi_j+\phi_{k_l}+\phi_{k_{\delta(l)}},& \mbox{$M$ odd},\\
   \end{matrix}
\right.,\nonumber\\
\eeaa
with $\delta(l)=M-(l-1)$.
If we choose in Eq.(\ref{fm2}) 
$\beta=-\pi(M-1)$, if $M$ is even, or $\beta=\pi(M-1)^2/2$, if $M$ is odd,  then 
we can choose $\xi_j=0$ 
[$j=1,\ldots,M$], to get
\beq
\label{rel-phases-base-1}
\theta_{k_l}+\theta_{k_{\delta(l)}}= \phi_{k_l}+\phi_{k_{\delta(l)}}.
\eeq
Notice that the phases $\phi_{k_l}$ can always be interpreted as the result of the mapping 
$|\psi_j\rangle \equiv e^{ih(\hat A)}|\tilde{\psi}_j\rangle$, with  $h$ being a real function, where $\phi_{k_l}=h(A_{k_l})$ and the states $|\tilde{\psi}_j\rangle$ of the projective measurement basis 
have the coefficients $\tilde{b}_{j,k_l}\equiv \langle \tilde{\psi}_j|\phi_+\rangle$, given in Eq.(\ref{mod-psi-j}),
with the phases $\tilde{\theta}_{j,k_l}= (j\pi/M)f_l+j\beta/M$. Therefore,
once we arbitrarily fix the phases $\theta_{k_l}$ of the initial state $\ket{\phi_+}$,  the states $\ket{\psi_j}$ of the projective measurement 
basis must have the phases $\theta_{j,k_l}$ given in 
Eq.(\ref{fase-psi-j}), with $\phi_{k_l}=h(A_{k_l})$ for any real function $h$.
This shows that the phases $\theta_{k_l}$ can be chosen arbitrarily, since the phases $\phi_{k_l}$
are arbitrary. 
Furthermore, we see that, for this example of projective measurement, there are no conditions on the real numbers $\xi_j$, so they can be chosen equal to zero.
\par
The family of projective measurements defined in Eq.(\ref{mod-psi-j}) and Eq.(\ref{fase-psi-j}) verify the balance condition in Eq. (\ref{balance-cond-bases})
regardless of the subset $\{\hat A_{k_l}\}_{l=1,\dots,M}$ of eigenstates of $\hat A$ present in the decomposition of the initial state $\ket{\phi_+}$. 
However,  Eq.(\ref{balance-cond-bases}) and the symmetry imposed by Eq.(\ref{media-bacana})
on the eigenvalues $\{A_{k_l}\}_{l=1,\ldots,M}$  guarantee that $ \bra{\psi_j}\hat{A}\ket{\psi_j}=
\langle\hat{A}\rangle_{+}$ for $j=1,\ldots,M$. 
\par
\subsubsection{Second type of projective measurements}
\par
The second example of a projective measurement basis $\{\ket{\psi_j}\}_{j=1,\ldots,M}$ that
allows a global saturation of the QIB is the one whose coefficients 
$b_{j,k_l}$ are given by
\bea
b_{j,k_l}&\equiv& \braket{A_{k_l}}{\psi_j}=\sqrt{\frac{(M-l)!(l-1)!}{(j-1)!(M-j)!}}\left(\frac{e^{i\vartheta}}{2}\right)^{l-\frac{M+1}{2}}\times\nonumber\\
&\times&P_{M-l}^{(l-j,l+j-(M+1))}(0),
\label{bj-ang-momentum}
\eea
where $P_{n}^{(\alpha,\beta)}(x)$ are the Jacobi polynomials \cite{Edmonds-book}, and $\vartheta$ is an arbitrary real number.   
These coefficients can be connected   to the matrix elements
\beq
d^{\rm j}_{m_z^\prime,m_z}(\pi/2)\equiv 
\mel{{\rm j},m_z^\prime}{e^{i\frac{\pi}{2\hbar}\hat J_y}}{{\rm j},m_z}
\label{matrix-d}
\eeq
in the theory of angular momentum \cite{Edmonds-book}, where $\ket{{\rm j},m_z}$ are eigenstates of the component $\hat J_z$ of the angular momentum operator $\hat{\bm{J}}$ , if the respective indexes are identified as  $M=2{\rm j}+1$, $l=m_z^\prime+(M+1)/2$ and
$j=m_z+(M+1)/2$. The condition $1 \leq l,j\leq M$  corresponds here to
the  constraint  $-{\rm j}\leq m_z^\prime,m_z\leq {\rm j}$. 
 Notice  that even values of $M$ correspond to half-integrals values of ${\rm j}$, 
 while odd values of $M$ correspond to integral values.
More specifically,  this mapping of indexes leads to the correspondence:
\beq
b_{j,k_l}\rightarrow e^{i(m_z^\prime \vartheta)}d^{\rm j}_{m_z^\prime,m_z}(\pi/2).
\eeq
\par 
Using the properties of the matrix elements $d^{\rm j}_{m_z^\prime,m_z}(\beta)$
 \cite{Edmonds-book}, it is easy to show that $|d^{\rm j}_{m_z^\prime,m_z}(\pi/2)|=
|d^{\rm j}_{-m_z^\prime,m_z}(\pi/2)|$, which is exactly the balance condition
$|b_{j,k_l}|=|b_{j,k_{\delta(l)}}|$, with $\delta(l)=M-(l-1)$. 
Since the real numbers $d^{\rm j}_{m_z^\prime,m_z}(\pi/2)$ are  elements of an orthogonal matrix
(real unitary matrix), the orthonormality of the states $\ket{\psi_j}$  is guaranteed.
Because the matrix elements $d^{\rm j}_{m_z^\prime,m_z}(\pi/2)$ are real numbers, 
we have for the phases of the coefficients $b_{j,k_l}$:
\beq
\theta_{j,k_l}=(l-(M+1)/2)\vartheta+s^{\prime\prime}_{l,j,M}\pi,
\label{theta-jota-particular}
\eeq 
where the integer $s^{\prime\prime}_{l,j,M}$ is such that 
$e^{is^{\prime\prime}_{l,j,M}\pi}=\sgn(P_{M-l}^{(l-j,l+j-(M+1))}(0))$.
Now, when $M>2$, using Eq.(\ref{ea3}), it is easy to see that, in this case,  the phases $\theta_{k_l}$ of the initial state $\ket{\phi_+}$
must satisfy:
\bse
\bea
&\theta_{k_l}+\theta_{k_{\delta(l)}}=0 \;\;\mod 2\pi
\label{phases-particular}\\
&(s^{\prime\prime}_{l,j,M}+s^{\prime\prime}_{\delta(l),j,M})\pi+\xi_j=0 \;\;\mod 2\pi.
\label{set-equa}
\eea
\ese
The set of Eqs.(\ref{set-equa}) determines the values of $\xi_j,\mod 2\pi$.
This implies that, in contrast to the use of the first family of  projective measurements,
 here the phases $\theta_{k_l}$ of the coefficients $c_{k_l}$, in the decomposition of the initial state $\ket{\phi_+}$ in the eigenbasis of the generator $\hat A$ (cf. Eq.(\ref{phi-base-ger})), are no longer completely arbitrary.  
\par 
Notice  that the subset $\{A_{k_l}\}_{l=1,\ldots,M}$
of eigenvalues of $\hat A$ that obey the symmetry in Eq.(\ref{ea1})
(see also Fig.\ref{fig2}), required
for  a global saturation of the QIB, are not necessarily  equally spaced. Thus, the states 
$\ket{\psi_j}=\sum_{j=1}^M b_{j,k_{l}}\ket{A_{k_l}}$, with the coefficients 
$b_{j,k_{l}}$ given in (\ref{bj-ang-momentum}),
are not necessarily 
equivalent to  eigenstates of an angular momentum operator.
However, when the eigenvalues $\{A_{k_l}\}_{l=1,\ldots,M}$ of
the operator $\hat A$ are equally spaced, the operator $\hat A$,  restricted to the subspace $\{\ket{A_{k_l}}\}_{l=1,\ldots,M}$,
is itself equivalent to an angular momentum  operator, and if we use the basis
$\{\ket{\psi_j}\}_{j=1,\ldots,M}$ with the coefficient $b_{j,k_l}$ given in 
(\ref{bj-ang-momentum}), then the states $\ket{\psi_j}$ are also 
eigenstates of an angular momentum  operator.
\par
\section{Some examples of global saturation of the QIB}
\label{SectionIV}
\par
The case in which the generator $\hat A$ is indeed an angular momentum component,
let's  say $\hat A=\hat J_z/\hbar$, was studied in \cite{Hofmann2009} in the context of 
phase estimation in  two path-interferometry,  using the Schwinger representation. 
In this case the parameter to be estimated, $x_v=\Delta \varphi_v$, is the phase difference between the 
 two paths. 
Our complete characterization of the structure of the initial states 
$\ket{\phi_+}$ and the projective measurements $\{\ket{\psi_j}\}$  that lead to a global saturation of the QIB contains the results presented in \cite{Hofmann2009} as special cases. Indeed, if we use  
 Eq.(\ref{cond-phi-psi-sat1}) together with  Eq.(\ref{phases-particular}),  we see that the initial states
that permit a global saturation of the QIB,  for phase estimation in two path-interferometry, satisfy:
\bea
\braket{{\rm j},m_z}{\phi_+}&=&|\braket{{\rm j},m_z}{\phi_+}|e^{i\theta_{m_z}}= \nonumber\\
&=&|\braket{{\rm j},-m_z}{\phi_+}|e^{-i\theta_{-m_z}}= \nonumber\\
&=&\braket{{\rm j},-m_z}{\phi_+}^*,
\label{cond-alemao}
\eea
with $-{\rm j}\leq m_z\leq {\rm j}$, $\theta_{m_z}\equiv\theta_{k_l}$, where the index $m_z$ is connected with $k_l=l$ by a suitable map.  
Eq.(\ref{cond-alemao}) is exactly the condition given in Eq.(8) of \cite{Hofmann2009} for  initial states $\ket{\phi_+}$ with a fixed  photon number $N=2{\rm j}$.
The projective measurement for a global saturation of the QIB in this case is 
$\{\ket{\psi_j}= \ket{{\rm j},m_x}\}$, where $\{\ket{{\rm j},m_x}\}$ are eigenstates of the 
$\hat J_x$ component of an angular momentum 
and the index $m_x$ is connected with $j$ by a suitable map. 
Notice that this is exactly the projective measurement  basis given by the coefficients 
in Eq.(\ref{matrix-d}), since $e^{i\frac{\pi}{2\hbar}\hat J_y}\ket{{\rm j},m_z}=\ket{{\rm j},m_x}$, and 
coincides with the projective measurement basis used in \cite{Hofmann2009}. 
The number $M_T$ of coefficients $\braket{{\rm j},m_z}{\phi_+}$  different from zero could be such that $M_T< M=2{\rm j}+1$, the total number of possible values of $m_z$. However, there is no difference for the saturation of the QIB
if we consider the subspace $\{\ket{{\rm j},m_z}\}$, with 
$m_z=l-(M+1)/2$ and $l=1,\ldots,M$, as the subspace where the initial state $\ket{\phi_+}$ lives. This subspace is equally spanned by the projective measurement $\{\ket{{\rm j},m_x}\}$, with $m_x=j-(M+1)/2$ and $j=1,\ldots,M$. 
\par
Our results show that all measurement basis of the family $\{e^{-i\varphi  \hat J_z/\hbar}\ket{{\rm j},m_x}\}$, where $\varphi$ is an arbitrary phase, lead to the saturation of QIB for the initial states that satisfy Eq.(\ref{cond-alemao}). That is, 
\bea
\mathcal{F}(\Delta \varphi_v,\{e^{-i\varphi \hat J_z/\hbar}\ket{{\rm j},m_x}\})&=&\mathcal{F}_Q\equiv4\langle(\Delta\hat{J}_z)^{2}\rangle_{+}=\nonumber\\
&=&4\langle \hat{J}_z^{2}\rangle_{+}
\eea 
for all values $\varphi$ (see Eq.(\ref{saturation-QIB})). Notice  that the initial states $\ket{\phi_+}$ that satisfy~(\ref{cond-alemao}) 
have $\langle \hat J_z \rangle_{+}=0$.
\par
The formalism used here assumes that the spectrum $\{A_{k_l}\}_{l=1,\ldots,M}$ 
corresponding to the subspace  $\{\ket{A_{k_l}}\}_{l=1,\ldots,M}$ where the initial state lives
is not degenerate.
This is not the case if the initial states has a fluctuating photon number,  {\it i.e.} 
\beq
\ket{\phi_+}=\sum_{\rm j}\sum_{m^{({\rm j})}_z} c_{m^{({\rm j})}_z}\ket{{\rm j},m^{({\rm j})}_z},
\label{phi-muitos-photons}
\eeq
with $c_{m^{({\rm j})}_z}\equiv \braket{{\rm j},m^{({\rm j})}_z}{\phi_+}$.
Since ${\rm j}=N/2$ is no longer fixed, eigenstates $\ket{{\rm j},m^{({\rm j})}_z}$ with equal values of $m^{({\rm j})}_z$ but different values of ${\rm j}$ could enter in the decomposition of $\ket{\phi_+}$. Such states, however, are eigentstates of $\hat J_z$ corresponding to the same eigenvalue $\hbar m^{({\rm j})}_z$. Nevertheless,   if the state in Eq.(\ref{phi-muitos-photons}) verifies the conditions in Eq.(\ref{cond-alemao})
for all values of ${\rm j}$, then it can be shown that  global saturation of the QIB 
can be reached via the 
projective measurement basis  
$\{\ket{\Psi_j}=\ket{{\rm j},m_x}\}$, with 
\beqq
\sum_{\rm j}\sum_{m^{({\rm j})}_x} \ket{{\rm j},m^{({\rm j})}_x}\bra{{\rm j},m^{({\rm j})}_x}=
\oplus_{{\rm j}=0}^{{\rm j}_{max}}\hat{\mathbb 1}_{\rm j},
\eeqq
where ${\rm j}_{max}$is  the largest value of ${\rm j}$ in the expansion in Eq.(\ref{phi-muitos-photons}).  However, one cannot  guarantee, in this case,  that those are the only states that permit  a global saturation of the QIB.
The coefficients $b_{j,k_l}\equiv \braket{A_{k_l}}{\Psi_j}$ are
\bea
b_{j,k_l}&\rightarrow&  \braket{{\rm j}^\prime,m^{({\rm j}^\prime)}_z}{{\rm j},m^{({\rm j})}_x} =\nonumber\\
&=&
\delta_{{\rm j}^\prime{\rm j}}\;e^{i(m_z^\prime \vartheta)}d^{\rm j}_{m_z^\prime,m_z}(\pi/2),
\eea
with ${\rm j}=0,\ldots,{\rm j}_{max}$ and $-{\rm j} \leq m_z^{({\rm j})}\leq {\rm j}$. 
Therefore, in each invariant subspace $\hat{\mathbb 1}_{\rm j}$, the corresponding 
coefficients $b_{j,k_l}$  are $e^{i(m_z^\prime \vartheta)}d^{\rm j}_{m_z^\prime,m_z}(\pi/2)$.
Notice that it is allowed to consider states with ${\rm j}_{max}\rightarrow \infty$.
\par 
It is interesting to show how  the global saturation of the QIB in the context of phase estimation with one bosonic
mode may happen. In this case, the generator $\hat A=\hat n=\hat a^\dagger \hat a$ is the 
number operator associated with the bosonic mode, described by the 
annihilation operator $\hat a$. Since the generator $\hat n$ has a non-degenerate 
spectrum, our results provide all the initial states $\ket{\phi_+}$
that allow a global saturation of the QIB under projective measurements.
Let's see how these states can be constructed.
Given a fixed  value for  $\langle \hat n \rangle_{+}$, since  the spectrum of $\hat n$ is equally spaced,  state $\ket{\phi_+}$ satisfies the 
symmetry condition in Eq.(\ref{ea1}) only if  $\langle \hat n \rangle_{+}$  coincides with some eigenvalue of $\hat n$ or is the arithmetic mean of any two eigenvalues. Then, all the eigenstates $\ket{n}$ with eigenvalues
$0 \le n\leq \langle \hat n \rangle_{+}$ and the eigenstates  symmetric to them 
with respect to the mean $\langle \hat n \rangle_{+}$, can be used to construct  an initial
state according to Eqs. (\ref{phi-base-ger}) and (\ref{cond-phi-psi-sat1}). It is, then, easy to se that because the spectrum of $\hat n$
is lower bounded, the number of terms 
in Eq.(\ref{phi-base-ger}) must be finite. 
This means, for example, that coherent states
\beqq
\ket{\phi_+}=\ket{\alpha}\equiv \sum_{n=0}^\infty e^{-|\alpha|^2/2} \frac{\alpha^{n}}{\sqrt{n!}} \ket{n},
\eeqq
with $\langle \hat n \rangle_{+}=|\alpha|^2$, are not among the initial states that 
allow a global saturation of the QIB under projective measurements. 
\par
However, if we consider coherent states with large values of 
$\langle \hat n \rangle_{+}=|\alpha|^2$, we can approximate 
the Poisson distribution by a Gaussian
\cite{Schleich-book}, {\it i.e.},
\beqq
p_n\equiv e^{-\langle \hat n \rangle_{+}} \frac{\langle \hat n \rangle_{+}^n}{n!} \approx
 \frac{e^{-\frac{1}{2\langle \hat n \rangle_{+}}\left(n-\langle \hat n \rangle_{+}\right)^2}}{\sqrt{2\pi \langle \hat n \rangle_{+} }}\equiv g_{n},
\eeqq
yielding 
\beq
\ket{\phi_+}=\ket{\alpha}\approx \sum_{n=0}^\infty \sqrt{g_n} e^{i\theta n}\ket{n}\approx
\sum_{n=0}^{M-1} \sqrt{g_n} e^{i\,n\,\theta}\ket{n}\;,
\label{coh-state-approx}
\eeq
with $M=2\langle \hat n \rangle_{+}+1$.
Clearly this state verifies the balance condition $\sqrt{g_n}=\sqrt{g_{2\langle \hat n \rangle_{+}+2-n}}$ in Eq.(\ref{cond-phi-psi-sat1}) so that it can saturate the QIB if we use the projective measurement basis
\beqq
\ket{\psi_j}=\frac{1}{\sqrt{M}}\sum_{n=0}^{M-1} e^{i\theta_{j,n}}\ket{n},
\eeqq
where the phases $\theta_{j,n}$ are given in Eqs.(\ref{fase-psi-j}) and (\ref{fm})
with $k_l=l=n-1$.
It is interesting to notice that,
because the phases in the state Eq.(\ref{coh-state-approx})
do not satisfy the conditions in Eq.(\ref{phases-particular}), it is not possible, in this case, to use the projective measurement basis defined in Eq.(\ref{bj-ang-momentum}). 

\section{Global saturation of the QIB and the Heisenberg Limit}
\label{SectionV}

A very relevant problem in quantum metrology consists in determining, for fixed resources, which are the states that reach  the largest possible QIB. Such states lead to the lowest possible Quantum Cramér-Rao bound, using those resources.
 For the pure state families given  in Eq.(\ref{our-families}), one can  consider  $\langle\hat A\rangle_+$ as the fixed resource. We show now that, for those families,  the largest QIB among all the initial states
$\ket{\phi_+}$ that allow a global saturation of that bound corresponds to
\bea
\mathcal{F}^{HL}_{Q}=4(\langle\hat{A}\rangle_{+}-A_{0})^{2},
\label{inf-q-fisher-M2}
\eea
when the generator $\hat A$ has a lower bounded spectrum. Here, $A_0$ is the lowest eigenvalue of $\hat A$.
The quantum Cramér-Rao bound $1/\nu\mathcal{F}^{HL}_{Q}$  is known in the literature 
as the Heisenberg limit \cite{Giovannetti2006}. This implies that the Heisenberg limit can be attained with projective measurements, without any previous information about the true value of the parameter and without the use of any adaptive estimation scheme. It also implies that the Heisenberg limit cannot be surpassed under these conditions.
\par
The initial states that permit a global saturation of the QIB and have a quantum Fisher information equal to $\mathcal{F}^{HL}_{Q}$ are written as:
\beaa
\ket{\phi^{HL}_+}=\frac{1}{\sqrt{2}}\left(\ket{A_{0}}+e^{i\theta_k}
\ket{A_k}\right),
\label{est-M2-sat-HL}
\eeaa
where $A_k\equiv 2\langle\hat A\rangle_+-A_0$ and $\theta_k$ is an arbitrary phase. 
The states in the projective measurements basis that lead to the saturation of the QIB, for the initial states above,  have the structure:
\beaa
\ket{\psi_1}&=&\frac{1}{\sqrt{2}}
\left(\ket{A_{0}}+
e^{i\theta_{1,k}}
\ket{A_k}\right),\\
\ket{\psi_2}&=&\frac{1}{\sqrt{2}}\left(\ket{A_{0}}-
e^{i\theta_{1,k}}
\ket{A_k}\right),
\eeaa
with $\theta_{1,k}$ an arbitrary phase.
\par
In order to show that $\mathcal{F}^{HL}_{Q}$ is the largest quantum information associated with the states that may globally saturate the QIB, notice that, for a fixed value of  $\langle\hat A\rangle_+$, there are several initial states $\ket{\phi_+^M}$ that can be decomposed in the form given in Eq.(\ref{phi-base-ger}), for $M\ge2$, which satisfy condition (\ref{cond-phi-psi-sat1}). All these states allow a global saturation of the QIB, that is
\beq
\mathcal{F}_M(x_v,\{\ket{\psi_j}\})=\mathcal{F}_Q^M=4\langle(\Delta\hat{A})^{2}\rangle_{\phi_+^M},
\eeq 
regardless of the value of $x_v$, where $\mathcal{F}_M(x_v,\{\ket{\psi_j}\})$ is the Fisher information associated with the projective measurement $\{\ket{\psi_j}\bra{\psi_j}\}$  on the states $\ket{\phi_+^M}$. Here, $\langle(\Delta\hat{A})^{2}\rangle_{\phi_+^M}=\Tr[\dyad{\phi_+^M}{\phi_+^M}(\hat A-\langle \hat A\rangle_{+})^2]$.
\par
Using condition (\ref{cond-phi-psi-sat1}), that is  $|c_{k_l}|=|c_{k_{\delta(l)}}|$,  we can write: 
\bea
&\mathcal{F}_Q^M\equiv4\langle(\Delta\hat{A})^{2}\rangle_{\phi_+^M}=8\sum_{l=1}^{\lceil M/2\rceil}|c_{k_l}|^{2}\left(\langle\hat{A}\rangle_{+}-A_{k_l}\right)^{2}=\nonumber\\
&=4\left(\langle\hat{A}\rangle_{+}-A_{k_1}\right)^{2}-4c\left(\langle\hat{A}\rangle_{+}-A_{k_1}\right)^{2}-\nonumber\\
&-8\sum_{l=2}^{\lceil M/2\rceil}|c_{k_l}|^{2}\left((\langle\hat{A}\rangle_{+}-A_{k_1})^{2}-(\langle\hat{A}\rangle_{+}-A_{k_l})^{2}\right)\leq\nonumber\\
&\leq 4\left(\langle\hat{A}\rangle_{+}-A_{k_1}\right)^{2}\equiv\mathcal{F}_Q^{M=2}\leq \nonumber\\
&\leq 4\left(\langle\hat{A}\rangle_{+}-A_{0}\right)^{2}\equiv\mathcal{F}_Q^{HL}
.
\label{var-phi-M>2}
\eea
Here, we  used $|c_{k_1}|^2=1/2-\sum_{l=2}^{s(M)}|c_{k_l}|^2-c/2$, where $s(M)$ is equal
to $\lceil M/2\rceil$ if $M$ is even and equal to
$\lceil M/2\rceil-1$ if $M$ is odd. We also set $c=0$ if $M$ is even and $c=|c_{k_{\lceil M/2\rceil}}|^2=\langle\hat A\rangle_+$ if $M$ is odd,
and we use that $|\langle\hat{A}\rangle_{+}-A_{0}|\geq|\langle\hat{A}\rangle_{+}-A_{k_1}|\geq |\langle\hat{A}\rangle_{+}-A_{k_l}|$, for  $l=2,\ldots,M$.
This shows that  $\mathcal{F}_Q^{HL}$ is the largest quantum Fisher information associated to the  initial states that allow a global saturation of the QIB. 
 
\section{Conclusion}
\label{SectionVI}
 In conclusion, we have considered the long standing quest to find all the initial states, together with the corresponding projective measurements, that allow a saturation of the Quantum Information Bound (QIB) without any previous information about the true value of the parameter to be estimated and without the use of any adaptive estimation scheme. We have been able to completely solve this problem for the important situation where information about the parameter is imprinted on an initial pure probe state via an unitary process whose generator does not depend explicitly on the parameter to be estimated. We have fully characterized all the initial states and corresponding projective measurements that allow a global saturation of the QIB under such conditions. We have also shown that, for a fixed mean value 
 $\langle\hat A\rangle_+$ of the generator of the unitary transformation, the largest quantum Fisher information associated to those states leads to the so-called Heisenberg limit. This implies that the Heisenberg limit can be attained with projective measurements, without any previous information about the true value of the parameter and without the use of any adaptive estimation scheme.


\begin{acknowledgements}
We acknowledge financial support from the Brazilian agencies FAPERJ, CNPq,  CAPES and the INCT-Informa\c{c}\~ao Qu\^antica. 
\end{acknowledgements}

\appendix

\section{}
\label{AppendixA}

Here we show that Eqs.(\ref{cond-sat-Im}) are satisfied if and only if the sets $\{A_{k_l}\}$,
$\{c_{k_l}\}$ and $\{b_{j,k_l}\}$ with $j,l=1,\ldots,M$, verify the conditions in 
Eqs.(\ref{cond-ea}), assuming that the set of eigenstates $\{\ket{A_{k_l}}\}_{l=1,\ldots,M}$  present in the decomposition  of $\ket{\phi_+}$ does not contain two eigenstates $\ket{A_{k_l}}$ corresponding to the same eigenvalue of $\hat A$. We start writing 
\begin{widetext}
\bea
&\mathrm{Im}\big[w_{j}(\epsilon)z^{*}_{j}(\epsilon)\big]=-\frac{2}{\sqrt{{\cal F}_Q}}
\sum_{l=1}^M\sum_{l^\prime=1}^M  |c_{k_l}|c_{k_{l^\prime}}|
|b_{j,k_l}||b_{j,k_{l^\prime}}| (A_{k_l}-\langle\hat A\rangle_+)
\cos\left((A_{k_{l^\prime}}-A_{k_l} )\epsilon+\theta_{k_l}-\theta_{j,k_l}-\theta_{k_{l^\prime}}+\theta_{j,k_{l^\prime}}\right)=\nonumber\\
&=
\sum_{l=1}^M\sum_{l^\prime=1}^M (A_{k_l}-\langle\hat A\rangle_+)\left(
h_{j,l,l^\prime}
\cos\left((A_{k_{l^\prime}}-A_{k_l} )\epsilon\right)-
g_{j,l,l^\prime}
\sin\left((A_{k_{l^\prime}}-A_{k_l} )\epsilon\right)\right),
\label{Eq1AppA}
\eea
\end{widetext}
where we use the expressions for $z_j(\epsilon)$ and $w_j(\epsilon)$ in Eqs.(\ref{def-w-and-z}). Furthermore, we use the identity $\cos(x+y)=\cos x\cos y-\sin x\sin y$ and  define 
\bse
\bea
h_{j,l,l^\prime}&\equiv&|c_{k_l}|c_{k_{l^\prime}}|
|b_{j,k_l}||b_{j,k_{l^\prime}}|\times\nonumber\\
&&\times\cos\left(\theta_{k_l}-\theta_{j,k_l}-\theta_{k_{l^\prime}}+\theta_{j,k_{l^\prime}}\right), \\ 
g_{j,l,l^\prime}&\equiv&|c_{k_l}|c_{k_{l^\prime}}|
|b_{j,k_l}||b_{j,k_{l^\prime}}|\times\nonumber\\
&&\times\sin\left(\theta_{k_l}-\theta_{j,k_l}-\theta_{k_{l^\prime}}+\theta_{j,k_{l^\prime}}\right).
\eea
\ese
Because the equality in Eqs.(\ref{Eq1AppA}) must hold for any value of $\epsilon$, we can
write those equations for  $-\epsilon$ and combine the two cases in order to arrive to the equivalent equations
\bse
\label{Eq-for-h-g}
\bea
&\sum_{l=1}^M h_{j,l,l}(A_{k_l}-\langle\hat A\rangle_+)+
\sum_{l=1_{l\neq l^\prime}}^M\sum_{l^\prime=1}^M\,h_{j,l,l^\prime}\times\nonumber\\
&\times(A_{k_l}-\langle\hat A\rangle_+)\cos\left((A_{k_{l^\prime}}-A_{k_l} )\epsilon\right)=0
\label{Eq-for-h}\\
&\sum_{l=1_{l\neq l^\prime}}^M\sum_{l^\prime=1}^M\,g_{j,l,l^\prime}\nonumber\\
&\times(A_{k_l}-\langle\hat A\rangle_+)\sin\left((A_{k_{l^\prime}}-A_{k_l} )\epsilon\right)=0,
\label{Eq-for-g}
\eea
\ese
that must be valid for all values of $\epsilon$ and  $j=1,\ldots,M$.
It is more convenient to rewrite Eqs.(\ref{Eq-for-h-g}) summing over indexes such  that $l< l^\prime$:
\bse
\bea
\sum_{l=1}^M\sum_{l^\prime=l+1}^M\,\tilde h_{j,l,l^\prime}
(\cos\left(\omega_{l^\prime l}\epsilon\right)-1)&=&0
\label{Eq-key-0}\\
\sum_{l=1}^M\sum_{l^\prime=l+1}^M\,\tilde g_{j,l,l^\prime}
\sin\left(\omega_{l^\prime l}\epsilon\right)&=&0,
\label{Eq-key-1}
\eea
\ese
where we define 
$\tilde g_{j,l,l^\prime}\equiv g_{j,l,l^\prime}
(A_{k_l}-\langle\hat A\rangle_++A_{k_{l^\prime}}-\langle\hat A\rangle_+)$ and  the frequencies $\omega_{l^\prime l}\equiv A_{k_{l^\prime}}-A_{k_l}$.
We also use that when we evaluate Eq.(\ref{Eq-for-h})  for  $\epsilon=0$ we have
$\sum_{l=1}^M h_{j,l,l}(A_{k_l}-\langle\hat A\rangle_+)=-\sum_{l=1_{l\neq l^\prime}}^M\sum_{l^\prime=1}^M\,h_{j,l,l^\prime}
(A_{k_l}-\langle\hat A\rangle_+)$.
\par
Note that, in principle, the frequencies $\omega_{l^\prime l}$ can be degenerate or non-degenerate. So, we can divide the sum in Eq.(\ref{Eq-key-1}) as 
\bse
\label{linear-indep-dos}
\bea
&\sum_l\sum_{l^\prime}\,\tilde g_{j,l,l^\prime}
\sin\left(\omega_{l^\prime l}\epsilon\right)+ \\
&\sum_{l^{\prime\prime}}\sum_{l{^{\prime\prime\prime}}}\,
\left[\sum_{l=l^{\prime\prime}}\sum_{l^\prime=l{^{\prime\prime\prime}}}\tilde g_{j,l,l^\prime}\right]
\sin\left(\omega_{l{^{\prime\prime\prime}}l^{\prime\prime}}\epsilon\right)=0,
\label{linear-indep}
\eea 
\ese
where the sums over the indexes $l,l^\prime$ correspond to the non-degenerate frequencies 
and the sums over the indexes $l^{\prime\prime},l^{\prime\prime\prime}$ over the degenerate ones. Note that we consider in Eq.(\ref{Eq-key-1}) that $l<l^\prime$, then $A_{k_l}<A_{k_{l^\prime}}$ because $k_1<\ldots< k_M$. Analogously, $l^{\prime\prime}<l^{\prime\prime\prime}$,
so that $A_{k_{l^{\prime\prime}}}<A_{k_{l^{\prime\prime\prime}}}$.
\par
Because the functions $\sin\left(\omega_{l^\prime l}\epsilon\right)$
and $\sin\left(\omega_{l{^{\prime\prime\prime}}l^{\prime\prime}}\epsilon\right)$ are linearly 
independent, the coefficient $\tilde g_{j,l,l^\prime}$ and $\sum_{l=l^{\prime\prime}}\sum_{l^\prime=l{^{\prime\prime\prime}}}\tilde g_{j,l,l^\prime}$, in Eqs.(\ref{linear-indep-dos}),
must be equal to zero for all values of $j$. Now, note that 
the coefficients of the expansion in Eq.(\ref{phi-base-ger}) of the initial 
state $\ket{\phi_+}$ are such that $c_{k_l}\neq 0$  for $l=1,\ldots,M$. Notice also that  the coefficient $b_{j,k_l}$ of the expansion of the measuring basis states 
$\ket{\psi_j}$, in Eq.(\ref{psij-na-base-de-A}), can only be zero for specific values of 
$l$ but not for all values of $j=1,\ldots,M$. So, from $\tilde g_{j,l,l^\prime}=0$, in 
Eq.(\ref{linear-indep}), we arrive to:
\beq
A_{k_l}-\langle\hat A\rangle_++A_{k_{l^\prime}}-\langle\hat A\rangle_+=0.
\label{first-cond}
\eeq
This condition simply says that, for non-degenerate frequencies 
$\omega_{l^\prime l}\equiv A_{k_{l^\prime}}-A_{k_l}$, the mean $\langle\hat A\rangle_+$
must be the arithmetic mean of the eigenvalues $A_{k_l}$   and $A_{k_{l^\prime}}$,
{\it i.e.} $(A_{k_{l^\prime}}+A_{k_l})/2=\langle\hat A\rangle_+$. 
Clearly, the frequency $\omega_{M,1}$
is non-degenerate because $\omega_{l^\prime l}<\omega_{M,1}$ for all values of 
$l,l^\prime=1,\ldots,M$ ($A_{k_1}<\ldots<A_{k_M}$). Therefore, we must have
$(A_{k_M}+A_{k_1})/2=\langle \hat A\rangle_+$, or equivalently,
\beq
\label{AkMAk1media}
A_{k_M}-\langle \hat A\rangle_++A_{k_1}-\langle \hat A\rangle_+=0.
\eeq
\par
Now, note that the frequencies $\omega_{M,2}$ and $\omega_{M-1,1}$ must be degenerate.  Otherwise, we would arrive to the contradictory  results
$\langle \hat A \rangle_+=(A_{k_M}+A_{k_2})/2>(A_{k_M}+A_{k_1})/2=\langle \hat A \rangle_+$ or  $\langle \hat A \rangle_+=(A_{k_{M-1}}+A_{k_1})/2<(A_{k_{M}}+A_{k_1})/2=\langle \hat A \rangle_+$ by  using Eq.(\ref{AkMAk1media}) and that $A_{k_2}>A_{k_1}$,  and
$A_{k_{M-1}}<A_{k_{M}}$. However, since $\omega_{l^\prime,l}<\omega_{M,2}$ and 
$\omega_{l^\prime,l}<\omega_{M-1,1}$ for all values of $l,l^\prime=2,\ldots,M$ and 
$\omega_{M,2},\omega_{M-1,1}<\omega_{M,1}$,
the only possibility is that $\omega_{M-1,1}=\omega_{M,2}$. This is equivalent to
the condition $(A_{k_{M-1}}+A_{k_2})/2=(A_{k_M}+A_{k_1})/2$, and, using Eq.(\ref{AkMAk1media}), it is also equivalent to 
\beq
\label{AkM-1Ak2media}
A_{k_{M-1}}-\langle \hat A\rangle_++A_{k_2}-\langle \hat A\rangle_+=0.
\eeq
We can now repeat the arguments for the frequencies $\omega_{M-1,3}$ and $\omega_{M-2,2}$. Indeed, this frequencies must be degenerate because, otherwise,
we arrive at the contradictory results
$\langle \hat A \rangle_+=(A_{k_{M-1}}+A_{k_3})/2>(A_{k_{M-1}}+A_{k_2})/2=\langle \hat A \rangle_+$ or  $\langle \hat A \rangle_+=(A_{k_{M-2}}+A_{k_2})/2<(A_{k_{M-1}}+A_{k_2})/2=\langle \hat A \rangle_+$ by using Eq.(\ref{AkM-1Ak2media}) and that $A_{k_3}>A_{k_2}$,
$A_{k_{M-2}}<A_{k_{M-1}}$. However, since $\omega_{l^\prime,l}<\omega_{M-1,3}$ and 
$\omega_{l^\prime,l}<\omega_{M-2,2}$ for all values of $l,l^\prime=3,\ldots,M$ and 
$\omega_{M-1,3},\omega_{M-2,2}<\omega_{M-1,2}$, the only possibility is that $\omega_{M-2,2}=\omega_{M-1,3}$. This is equivalent to
the condition $(A_{k_{M-2}}+A_{k_3})/2=(A_{k_{M-1}}+A_{k_2})/2$, and, using Eq.(\ref{AkM-1Ak2media}), it is also equivalent to 
\beq
\label{AkM-2Ak3media}
A_{k_{M-2}}-\langle \hat A\rangle_++A_{k_3}-\langle \hat A\rangle_+=0.
\eeq
These  two steps  illustrate the iterative process to be followed.
They show that the frequencies $\omega_{\delta(l),1+l}$ and $\omega_{M-l,l}$, with $\delta(l)=M-(l-1)$ and 
$l=1,\ldots,s(M)$, where $s(M)\equiv \lceil M/2 \rceil$ if $M$ is even and $s(M)\equiv \lceil M/2 \rceil-1$ if
$M$ is odd, are degenerate in such a way that
$\omega_{\delta(l),1+l}=\omega_{M-l,l}$ and that they are different from any other 
frequencies. This is enough to prove that 
$A_{k_{\delta(l)}}-\langle \hat A\rangle_++A_{k_l}-\langle \hat A\rangle_+=0$,
that is exactly the condition in Eq.(\ref{ea1}). This symmetry condition for the spectrum of 
eigenvalues $\{A_{k_l}\}_{l=1,\ldots,M}$, of $\hat A$, that enter in the decomposition 
of the initial state $\ket{\phi_+}$, is illustrated in Fig.\ref{fig1}, where we see that when
$M$ is odd necessarily $A_{\lceil M/2\rceil}=\langle \hat A\rangle_+$.
\par
Because $\omega_{\delta(l),1+l}=\omega_{M-l,l}$ for $l=1,\ldots,s(M)$, and because these frequencies are different from any other frequencies, the coefficient of 
$\sin(\omega_{\delta(l),1+l})=\sin(\omega_{M-l,l})$ in Eq.(\ref{linear-indep}) must be equal to zero, {\it i.e.} $(g_{j,\delta(l),l+1}+g_{j,l,M-l})(A_{k_{\delta_l}}-\langle\hat A\rangle_++
A_{k_{l+1}}-\langle\hat A\rangle_+)=0$.
Analogously, the coefficient of 
$\cos(\omega_{\delta(l),1+l})-1=\cos(\omega_{M-l,l})-1$ in Eq.(\ref{Eq-key-0}) must be equal to zero, {\it i.e.} $(h_{j,\delta(l),l+1}-h_{j,l,M-l})(A_{k_{\delta_l}}-\langle\hat A\rangle_++
A_{k_{l+1}}-\langle\hat A\rangle_+)=0$.
This leads to the following sets of equations:
\begin{widetext}
\bse
\label{e10}
\bea
&|c_{k_{\delta(l)}}| |c_{k_{l+1}}| |b_{j,k_{\delta(l)}}| |b_{j,k_{l+1}}| 
\sin(\theta_{k_{\delta(l)}}-\theta_{j,k_{\delta(l)}}-\theta_{k_{l+1}}+\theta_{j,k_{l+1}})=\nonumber\\
&=-|c_{k_{M-l}}| |c_{k_l}| |b_{j,k_{M-l}}| |b_{j,k_{l}}| 
\sin(\theta_{k_{l}}-\theta_{j,k_l}-\theta_{k_{M-l}}+\theta_{j,k_{M-l}}),\label{e10a}\\
&|c_{k_{\delta(l)}}| |c_{k_{l+1}}| |b_{j,k_{\delta(l)}}| |b_{j,k_{l+1}}| 
\cos(\theta_{k_{\delta(l)}}-\theta_{j,k_{\delta(l)}}-\theta_{k_{l+1}}+\theta_{j,k_{l+1}})=
\nonumber\\
&=|c_{k_{M-l}}| |c_{k_l}| |b_{j,k_{M-l}}| |b_{j,k_{l}}|
\cos(\theta_{k_{l}}-\theta_{j,k_l}-\theta_{k_{M-l}}+\theta_{j,k_{M-l}})\label{e10b}.
\eea
\ese
\end{widetext}
By taking the square in both Eqs.(\ref{e10a}) and (\ref{e10b}) and adding  them, we get
\bea
|c_{k_{\delta(l)}}|  |b_{j,k_{\delta(l)}}| |c_{k_{l+1}}| |b_{j,k_{l+1}}|=\nonumber\\
=|c_{k_{\delta(l+1)}}| |b_{j,k_{\delta(l+1)}}| |c_{k_l}| |b_{j,k_{l}}|, 
\label{e11ap}
\eea
where $l=1,2,\ldots,s(M)$ and we use that 
$\delta(l+1)=M-l$. 
Applying $l=s(M)$ in Eq.(\ref{e11ap}), and noting that 
if $M$ is even $\delta(s(M)+1)=s(M)$ and $\delta(s(M))=s(M)+1$ and if 
$M$ is odd $\delta(s(M)+1)=s(M)+1$, we arrive at
\bea
|c_{k_{s(M)}}| |b_{j,k_{s(M)}}|=|c_{k_{\delta(s(M))}}| |b_{j,k_{\delta(s(M))}}|.
\label{e12ap}
\eea
After that, we substitute $l=s(M)-1$ in Eq.(\ref{e11ap}). In the resulting expression, we apply Eq.(\ref{e12ap}) in order to have
\bea
|c_{k_{s(M)-1}}| |b_{j,k_{s(M)-1}}|=|c_{k_{\delta(s(M)-1)}}| |b_{j,k_{\delta(s(M)-1)}}|.
\label{e13ap}
\eea
Then, we use $l=s(M)-2$, together with Eq.(\ref{e13ap}), in Eq.(\ref{e11ap}). Following this iterative procedure, we are able to show that
\bea
|c_{k_{l}}| |b_{j,k_{l}}| = |c_{k_{\delta(l)}}| |b_{j,k_{\delta(l)}}|,
\label{e14ap}
\eea
for $l=1,2,\ldots,s(M)$, which is the result of the Eq.(\ref{ea2}).
\par
Now, we plug Eq.(\ref{e14ap}) into Eqs.(\ref{e10}) and, by solving the resulting  system of equations, we get to the following solution
\bea
(\theta_{k_{\delta(l)}}-\theta_{j,k_{\delta(l)}})+(\theta_{k_{l}}-\theta_{j,k_l})=\nonumber\\
=(\theta_{k_{l+1}}-\theta_{j,k_{l+1}})+(\theta_{k_{M-l}}-\theta_{j,k_{M-l}}).
\label{e15ap}
\eea
If we apply $l=s(M)-1$ in Eq.(\ref{e15ap}) ($l=s(M)$ does not give any extra information about the phase relation), we have
\bea
(\theta_{k_{s(M)-1}}-\theta_{j,k_{s(M)-1}}) + (\theta_{k_{\delta(s(M)-1)}}-\theta_{j,k_{\delta(s(M)-1)}})=\nonumber\\
=(\theta_{k_{s(M)}}-\theta_{j,k_{s(M)}}) + (\theta_{k_{\delta(s(M))}}-\theta_{j,k_{\delta(s(M))}}).\nonumber\\
\label{e16ap}
\eea
If we use $l=s(M)-2$ in Eq.(\ref{e15ap}) and then substitute Eq.(\ref{e16ap}) in the result, we obtain :
\bea
(\theta_{k_{s(M)-2}}-\theta_{j,k_{s(M)-2}}) + (\theta_{k_{\delta(s(M)-2)}}-\theta_{j,k_{\delta(s(M)-2)}})=\nonumber\\
=(\theta_{k_{s(M)}}-\theta_{j,k_{s(M)}}) + (\theta_{k_{\delta(s(M))}}-\theta_{j,k_{\delta(s(M))}}).\nonumber\\
\label{e17ap}
\eea
If we put $l=s(M)-3$ in Eq.(\ref{e15ap}) and plug Eq.(\ref{e17ap}) into the result, we will find the same term of the right-hand side of the Eqs.(\ref{e16ap}) and (\ref{e17ap}). Repeating these steps iteratively for all the remaining terms, we will see that the terms $(\theta_{k_{\delta(l)}}-\theta_{j,k_{\delta(l)}})+(\theta_{k_{l}}-\theta_{j,k_l})$ are equal for all $l=1,2,\ldots,s(M)$. So, as in principle the phases are all different from each other, this equality among all the expressions only holds if 
\bea
(\theta_{k_{\delta(l)}}-\theta_{j,k_{\delta(l)}})+(\theta_{k_{l}}-\theta_{j,k_l})=\xi_{j},
\label{e18ap}
\eea
where $\xi_{j}$ is a constant depending only on $j$. The Eq.(\ref{e18ap}) is that one in Eq.(\ref{ea3}).

\section{}
\label{AppendixB}

Here we prove that one solution to the conditions 
\beq
\label{diag-iguales}
\bra{\psi_j}\hat{A}\ket{\psi_j}=\alpha \;\;,\;\; j=1,\ldots,M,
\eeq
is given by states $\ket{\psi_j}$ whose expansion in the eigenbasis of the generator $\hat A$ is of the form shown in Eq.(\ref{psij-na-base-de-A}), with
the coefficients given in
Eq.(\ref{mod-psi-j}) and the phases in (\ref{fase-psi-j}) and (\ref{fm}).
Remember that, since all the coefficients $b_{j,k_l}\neq 0$, the 
subspace spanned by
$\{\ket{\psi_j}\}_{j=1,\ldots,M}$ and the subspace spanned by$\{\ket{A_{k_l}}\}_{l=1,\ldots,M}$ 
coincide.
Let's start defining an auxiliary unitary operator $\hat V$ within the subspace 
$\{\ket{\psi_j}\}_{j=1,\ldots,M}$ of the system Hilbert space, such that
\bea
\hat{V}\ket{\psi_j}&=&\ket{\psi_{j+1}},\label{e1.1}\\
\hat{V}\ket{\psi_M}&=&e^{i\beta}\ket{\psi_{1}},
\label{e1.2}
\eea
where $1\leq j\leq (M-1)$ and $\beta$ is an arbitrary phase.
We call $\hat V$ the {\it shift}  operator over the basis $\{\ket\psi_j\}_{j=1,\ldots,M}$. It is important to notice that, for every 
finite basis, it is always possibly to define an operator $\hat V$ that shift the elements
of the basis. 
The unitary matrix, in the basis  $\{\ket{\psi_j}\}_{j=1,\ldots,M}$,  that  represent $\hat V$ is
\[
\left(\begin{array}{ccccccc} 0 & 0 & 0 & 0 & \cdots & 0 & e^{i\beta} \\ 
1 & 0 & 0 & 0 & \cdots & 0 &  0 \\
0 & 1 & 0 & 0 & \cdots & 0 & 0 \\
0 & 0 & 1 & 0 &  \cdots & 0 & 0 \\
 \vdots & \vdots &\vdots &\vdots &\ddots & \vdots& \vdots  \\ 
 0 & 0 & 0 & 0 & \cdots & 1 & 0
\end{array}\right).\]
Its eigenvalues are of the form $e^{i d_l}=e^{i(f_{l}\pi+\beta)/M}$ and
the eigenvectors of the form
\bea
\ket{d_{l}}=\frac{1}{\sqrt{M}}\sum_{j=1}^{M}e^{-i\theta^\prime_{j,l}}\ket{\psi_{j}},
\label{cm-base-ger}
\eea
with
\bea
\theta^\prime_{j,l}=(j\pi/M)\,f_l+j\beta/M,
\label{fase-cm}
\eea
and $f_l$ given in Eq.(\ref{fm}).
Therefore, the subspace $\{\ket{\psi_j}\}$ can be equivalently described by 
the basis $\{\ket{d_l}\}_{l=1,\ldots,M}$ formed by the eigenstates 
of the {\it shift} operator. We emphasise here that the states 
$\ket{d_l}$ belong to the system Hilbert space, which could have an arbitrary 
dimension. The matrix whose elements are 
\beq
\braket{\psi_j}{d_l}=(1/\sqrt{M})e^{-i\theta^\prime_{j,l}}
\label{unitary-matrix}
\eeq 
is unitary, so we can invert the relation in Eq.(\ref{cm-base-ger}) to write:
\beq
\ket{\psi_j}=\frac{1}{\sqrt{M}}\sum_{l=1}^M\;e^{i\theta^\prime_{j,l}}\ket{d_l}.
\label{psij-na-base-shift}
\eeq
\par
We can express the unitary {\it shift}  operator as $\hat V=e^{i\hat D}$, where 
$\hat D$ is a Hermitian operator with eigenvalues 
\beq
d_l\equiv(f_{l}\pi+\beta)/M,
\eeq 
and eigenvectors given in Eq.(\ref{cm-base-ger}). Notice that $\hat D$ has a non-degenerate 
spectrum and  that its diagonal elements  in the basis $\{\ket{\psi_j}\}$
are all equals, {\it i.e.}
\bea
\bra{\psi_{j}}\hat{D}\ket{\psi_j}=\frac{1}{M}\sum_{l=1}^{M}d_{l}\equiv \alpha_{1},
\label{e11}
\eea
where $\alpha_1$ does not depend on the value of ``$j$''.
\par
Now, remember that we are looking for the states $\ket{\psi_j}$
that verify~(\ref{diag-iguales}). This is a similar condition to the one in (\ref{e11})
for the generator $\hat D$ of the shift operator $\hat V=e^{i\hat D}$.
Let's us show that it is possible to consider that $\hat A$ is diagonal 
in the subspace spanned by the basis $\{\ket{d_l}\}_{l=1,\ldots,M}$ of eigenstates of $\hat D$. 
We first note that
\bea
\bra{\psi_{j=M}}d_{l}\rangle=\frac{e^{-i\beta}}{\sqrt{M}},
\label{e5}
\eea
for all values $l=1,\ldots,M$. 
Addittionaly, using Eq.(\ref{fm})
 in Eq.(\ref{fase-cm})  we get for $l\neq l^\prime$ the phases differences:
\bea
&\theta^\prime_{j,l}-\theta^\prime_{j,l^\prime}=\label{dif-fase-1}\\
&=\left\{
   \begin{matrix} 
      \frac{j\pi}{M}\left((l-l^\prime)(M+1)+[(-1)^{l}-(-1)^{l^\prime}]/2\right),\\
       \;\mbox{for $M$ even},\; \\
      \frac{j\pi}{M}(l-l^\prime)(M+1), \; \mbox{for $M$ odd}.\\
   \end{matrix}
\right.\nonumber
\eea
In particular, we have:
\bse
\bea
&\theta^\prime_{j=M,l}-\theta^\prime_{j=M,l^\prime}=2n\pi\\   
&\theta^\prime_{j,l}-\theta^\prime_{j,l^\prime}\neq 2n\pi\;\;\mbox{for $1\leq j\leq (M-1)$}, 
\label{diff-phases-not-zero}
\eea
\ese
with $n$ an integer. 
Now, we obtain
\bea
\bra{\psi_{j=M}}\hat{A}\ket{\psi_{j=M}}&=&\frac{1}{M}\bigg\{\sum_{l=1}^{M}\bra{d_{l}}\hat{A}\ket{d_{l}}+\nonumber\\
&+&\sum_{l\neq l^\prime}^{M}\bra{d_{l}}\hat{A}\ket{d_{l^\prime}}\bigg\}=\nonumber\\
&=&\alpha.
\label{e15}\\
\bra{\psi_{j\neq M}}\hat{A}\ket{\psi_{j\neq M}}&=&\frac{1}{M}\bigg\{\sum_{l=1}^{M}\bra{d_{l}}\hat{A}\ket{d_{l}}+\nonumber\\
&+&\sum_{l\neq l^\prime}^{M}e^{i(\theta^\prime_{j,l}-\theta^\prime_{j,l^\prime})}\bra{d_{l}}\hat{A}\ket{d_{l^\prime}}\bigg\}\nonumber\\
&=&\alpha.
\label{e16}
\eea
Since $(\theta^\prime_{j,m}-\theta^\prime_{j,m'})\neq2n\pi$ (see Eq.(\ref{diff-phases-not-zero})),
comparing Eq.(\ref{e15}) with Eq.(\ref{e16}), we see that one possibility is that
\bea
\bra{d_{l}}\hat{A}\ket{d_{l'}}=0,
\label{e17}
\eea
for all $l\neq l'$, which means that $\hat A$ is diagonal in the subspace spanned by 
$\{\ket{d_l}\}_{l=1,\ldots,M}$. 
The other possibility is that the second terms in Eqs. (\ref{e15}) and (\ref{e16}) are null. 
It is interesting to notice that this second possibility is verified if we use the coefficients 
in Eq.(\ref{bj-ang-momentum}) to define the states $\ket{\psi_j}$ through~(\ref{psij-na-base-de-A}) and then  use those states in the definition of the eigenstates of the {\it shift} operator in~(\ref{cm-base-ger}).    
\par 
Because the operator $\hat D$ is non-degenerate, the result in Eq.(\ref{e17}) means that 
we can identify the eigenstates of $\hat D$ and the eigenstates of $\hat A$ in the subspace $\{\ket{\psi_j}\}_{j=1,\ldots,M}$.
The order of this identification is unimportant, so we can set 
\beq
\label{direct-association}
\ket{d_l}=\ket{A_{k_l}}\;,\;l=1,\ldots,M.
\eeq 
\par
In order to obtain the projective measurement with states given in  Eq.(\ref{psij-na-base-de-A}),
with coefficient given in (\ref{mod-psi-j}) whose phases are given in (\ref{fase-psi-j}), we observe that if we apply an arbitrary unitary evolution $e^{i(h(\hat{A}))}$ 
to the states in Eq.(\ref{psij-na-base-shift}) (here $h(\hat{A})$ is any Hermitian operator
that depends on $\hat A$),
we obtain an equivalently admissible projective measurement [one that also fulfil the condition
that  all the matrix elements $\bra{\psi_j}\hat A\ket{\psi_j}$ are equal].
This is the reason why we include the extra phases $\phi_{k_l}\equiv h(A_{k_l})$ in Eq.(\ref{fase-psi-j}) in comparison with the phases in Eq.(\ref{fase-cm}).  
\par
We can verify the consistency of our results looking at 
the orthonormality relation:
\bea
\bra{\psi_j}\psi_{j^\prime}\rangle&=&\frac{1}{M}\sum_{l=1}^M e^{i(\theta_{j,k_l}-\theta_{j^\prime,k_l})}=\nonumber\\
&=&e^{i(\gamma_{j}-\gamma_{j^\prime})}e^{i(j-j^\prime)\beta/M}
\frac{1}{M}
\sum_{l=1}^M
e^{i\pi(j-j^\prime)f_l/M}=\nonumber\\
&=&\delta_{jj^\prime},
\label{ortonormalidade}
\eea
where we use 
that 
\beq
\label{deltajj}
\frac{1}{M}\sum_{l=1}^M
e^{i\pi(j-j^\prime)f_l/M}=\delta_{jj^\prime}.
\eeq
For $j=j^\prime $ we can immediately check this equality. 
In order to check the equality in Eq.(\ref{deltajj}) for  $j\neq j^\prime$, 
we proceed as follows. For $M$ even, we have :
\begin{widetext}
\bea
\sum_{l=1}^{M}e^{i\pi(j-j^{\prime})f_{l}/M}&=&\sum_{l\rightarrow\textit{even}}^{M}e^{i\pi(j-j^{\prime})(l+M-2)/M}+\sum_{l\rightarrow\textit{odd}}^{M}e^{i\pi(j-j^{\prime})(l-1)/M}=\nonumber\\
&=&e^{i\pi(j-j^{\prime})}\left(1+e^{2i\pi(j-j^{\prime})/M}+e^{4i\pi(j-j^{\prime})/M} + \ldots \right)+\left(1+e^{2i\pi(j-j^{\prime})/M}+e^{4i\pi(j-j^{\prime})/M} + \ldots \right)=\nonumber\\
&=&\frac{(1+e^{i\pi(j-j^{\prime})})(1-e^{i\pi(j-j^{\prime})})}{1+e^{2i\pi(j-j^{\prime})/M})}=\frac{(1-e^{2i\pi(j-j^{\prime})})}{1+e^{2i\pi(j-j^{\prime})/M})}=0,
\eea
\end{widetext}
and for $M$ odd, we have
\bea
&&\sum_{l=1}^{M}e^{i\pi(j-j^{\prime})f_{l}/M}=\sum_{l=1}^{M}e^{i\pi(j-j^{\prime})(l-1)(1-M)/M}=\nonumber\\
&=&1+e^{i\pi(j-j^{\prime})(1-M)/M}+e^{2i\pi(j-j^{\prime})(1-M)/M}+\ldots=\nonumber\\
&=&\frac{1-e^{-i\pi(j-j^{\prime})(M-1)}}{1+e^{i\pi(j-j^{\prime})(1-M)/M}}=0,
\eea
since $M-1$ is an even number.

\bibliographystyle{apsrev}

\end{document}